\documentclass[aoas,MSNbibl,nameyear,dvips]{arximspdf}
\usepackage{dcolumn,rotating}
\usepackage{graphicx}

\doi{10.1214/13-AOAS675} %kopijuoti is PTS
\volume{7}
\issue{4}
\pubyear{2013}
\firstpage{2361}
\lastpage{2383}

\makeatletter
\newcolumntype{d}[1]{D{.}{.}{#1}}
\newcommand{\eqref}[1]{(\ref{#1})}
\newcommand{\E}{\mathrm{E}}

\newcommand{\bs}[1]{\bolds{#1}}
\makeatother

\begin{document}
\begin{frontmatter}

\title{A semiparametric approach to mixed outcome latent variable models:
Estimating the association between cognition and regional brain
volumes\thanksref{T1}}
\runtitle{A semiparametric latent variable model}
\thankstext{T1}{Supported in part by Grant R01 AG 029672 from the
National Institute on Aging.
Data collection was supported by Grant P01 AG12435.}

\begin{aug}
\author[A]{\fnms{Jonathan}~\snm{Gruhl}\corref{}\ead[label=e1]{gruhlj@stat.washington.edu}\thanksref{tt1}},
\author[A]{\fnms{Elena A.} \snm{Erosheva}\ead[label=e2]{elena@stat.washington.edu}\thanksref{tt1}}
\and
\author[B]{\fnms{Paul K.} \snm{Crane}\ead[label=e3]{pcrane@u.washington.edu}\thanksref{tt2}}
\runauthor{J. Gruhl, E.~A. Erosheva and P.~K. Crane}
\affiliation{University of Washington\thanksmark{tt1} and Harborview
Medical Center\thanksmark{tt2}}
\address[A]{J. Gruhl\\
E. A. Erosheva\\
Department Of Statistics\\
University of Washington\\
Seattle, Washington 98195-4322\\
USA\\
\printead{e1}\\
\phantom{E-mail:\ }\printead*{e2}}

\address[B]{P. K. Crane \\
Department of Medicine \\
Harborview Medical Center\\
325 Ninth Avenue \\
Campus Box 359780 \\
Seattle, WA 98104\\
USA\\
\printead{e3}}
\end{aug}

% HISTORY:
\received{\smonth{3} \syear{2012}}
\revised{\smonth{7} \syear{2013}}

% ABSTRACT
%
\begin{abstract}
Multivariate data that combine binary, categorical, count and
continuous outcomes are common in the social and health sciences.
We propose a semiparametric Bayesian latent variable model for
multivariate data of arbitrary type that does not require specification
of conditional distributions. Drawing on the extended rank likelihood
method by Hoff [\textit{Ann. Appl. Stat.} \textbf{1} (2007) 265--283], we develop a semiparametric
approach for latent variable modeling with mixed outcomes and propose
associated Markov chain Monte Carlo estimation methods. Motivated by
cognitive testing data, we focus on bifactor models, a special case of
factor analysis. We employ our semiparametric Bayesian latent variable
model to investigate the association between cognitive outcomes and
MRI-measured regional brain volumes.
\end{abstract}
%
% KEYWORDS
% Pirmas kwd is didziosios raides
%
\begin{keyword}
\kwd{Latent variable model}
\kwd{Bayesian hierarchical model}
\kwd{extended rank likelihood}
\kwd{cognitive outcomes}
\end{keyword}

\end{frontmatter}

%s1 #&#
\section{Introduction}

Multivariate outcomes are common in medical and social studies. Latent
variable models provide means for studying the interdependencies among
multiple outcomes perceived as measures of a common concept or
concepts. These outcomes in many cases may be of mixed types in the
sense that some may be binary, others may be counts, and yet others may
be continuous. Most common latent variable models, however, have been
developed for outcomes of the same type. For example, standard factor
analysis models [\citet{bartholomew2011latent}] assume normally
distributed outcomes, item response theory models [\citet
{van1997handbook}] are typically applied to binary responses, and
graded response [\citet{samejima1969estimation}] and generalized partial
credit models [\citet{muraki1992generalized}] have been developed
specifically for ordered categorical data.

Existing research on latent variable models for mixed outcomes is
largely focused on two parametric approaches. The first approach is to
specify a different generalized linear model for each outcome that best
suits its type (e.g., binary, count, ordered categorical) and to
include shared latent variables as predictors that induce dependence
among the outcomes. \citet{sammel1997latent} and \citet
{moustaki2000generalized} developed this approach, referred to as
generalized latent trait modeling, employing the EM algorithm for
estimation. In a Bayesian framework, \citet{dunson2003dynamic} extended
the generalized latent trait models to allow for repeated measurements,
serial correlations in the latent variables and individual-specific
response behavior. The second approach to analyzing mixed discrete and
continuous outcomes with latent variables is the underlying latent
response approach where observed mixed outcomes are assumed to have
underlying latent responses that are continuous and normally
distributed. Introduction of the continuous latent responses enables
one to proceed with the analysis as one might for any multivariate
normal data. To map the underlying latent responses to observed mixed
outcomes, one must estimate threshold parameters. In this context,
\citet
{shi1998bayesian} employed Bayesian estimation for factor analysis with
polytomous and continuous outcomes. However, as noted by \citet
{dunson2003dynamic}, the underlying latent response approach is limited
in that some observed outcome types such as counts may not be easily
linked to underlying continuous responses.

Generalized latent trait models can be extended to account for
additional types of outcomes [\citet{skrondal2004generalized}],
including censored and duration outcomes. However,
%as we experienced in our analysis of cognitive testing data from the
%Subcortical Ischemic Vascular Dementia (SIVD) study
accommodating many possible types of outcomes that one may encounter in
practice may be time-consuming, susceptible to misspecification and of
little interest in its own right.

Our motivating example is a data set from a large multicenter study
called the Subcortical Ischemic Vascular Dementia (SIVD) Program
Project Grant [\citet{chui2006cognitive}]. The SIVD study collects
serial imaging and neuropsychological data from a large group of study
participants. One major study goal is to further elucidate
relationships between brain structure (as measured by MRI imaging) and
function (as measured by performance on neuropsychological tests). In
particular, investigators were especially interested in cerebrovascular
disease as manifested on MRI. Thus, we focus our analysis on one
particular cognitive domain, namely, executive functioning, that is
thought to be particularly susceptible to cerebrovascular disease
[\citet
{hachinski2006national}].
Executive functioning refers to higher order cognitive tasks
(``executive'' tasks) such as working memory, set shifting, inhibition
and other frontal lobe-mediated functions.
The SIVD study follows individuals longitudinally until death,
collecting results from repeated neuropsychological testing and brain
imaging. In this paper, we are concerned with relating individual
levels of executive functioning at the initial SIVD study visit to the
concurrent MRI-measured\vadjust{\goodbreak} amount of white matter hyperintensities (WMH)
located in the frontal lobe of the brain. Executive functioning
capabilities may be particularly sensitive to white matter
hyperintensities in this region [\citet{kuczynski2010white}].

The SIVD neuropsychological battery includes 21 distinct indicators
that can be conceptualized as measuring some facet of executive
functioning. We refer to the executive functioning-related outcomes as
``indicators,'' as they include some elements that are scales by
themselves and other elements that are subsets of scales.
Observed responses to the SIVD neuropsychological tests items are
diverse in their types.
In addition to binary and ordered categorical outcomes, the SIVD
neuropsychological indicators include count as well as censored count data.

In this paper, we develop a semiparametric approach to mixed outcome
latent variable models that avoids specification of outcome conditional
distributions given the latent variables.
Following the extended rank likelihood approach of \citet
{hoff2007extending}, we start by assuming the existence of continuous
latent responses underlying each observed outcome. We then make use of
the fact that the ordering of the underlying latent responses is
assumed to be consistent with the ordering of the observed outcomes.
This approach is similar to that of \citet{shi1998bayesian} but does not
require estimating unknown thresholds. When the data are continuous,
our approach is analogous to the use of a rank likelihood [\citet
{pettitt1982inference}]. When the data are discrete, our approach
relies on the assumption that the ordering of the latent responses is
consistent with the partial ordering of the observed outcomes. \citet
{hoff2007extending} introduced this general approach for estimating
parameters of a semiparametric Gaussian copula model with arbitrary
marginal distributions and designated the resulting likelihood as the
\textit{extended rank likelihood}. \citet{dobra2011copula} applied the
extended rank likelihood methods to the estimation of graphical models
for multivariate mixed outcomes.

Motivated by SIVD cognitive testing data, we specify a bifactor latent
structure for the semiparametric latent variable model. The bifactor
structure assumes existence of a general factor and some secondary
factors that account for residual dependency among groups of items
[\citet{holzinger1937bi,reise2007role}]. The bifactor model is a useful
tool for modeling the neuropsychological battery used in the SIVD
study, as it retains a single underlying executive functioning factor
while accounting for local dependencies among groups of related items.
The original idea for this work was presented earlier by
[\citeauthor{gruhl2010analyzing} (\citeyear{gruhl2010analyzing,gruhl2011semiparametric})]. \citet
{murray2011bayesian} recently proposed a closely related factor
analytic model for mixed data.

The remainder of this paper is organized as follows. We review the
semiparametric Gaussian copula model and introduce the new
semiparametric latent variable model in Section~\ref{sec:SPLVM-model}.
We develop Bayesian estimation approaches for the semiparametric latent
variable model in Section~\ref{sec:SPLVM-estimation}.
We extend the model hierarchically to include covariates in
Section~\ref{sec:SPLVM-hierarchical}. In Section~\ref{sec:SIVD-app} we briefly
demonstrate the performance of the model using simulated data before
focusing on the analysis of the SIVD data.

%s2 #&#
\section{Semiparametric latent variable model for mixed
outcomes}\label
{sec:SPLVM-model}

%s2.1 #&#
\subsection{Model formulation}

Let $i = 1, \ldots, I$ denote the $i$th participant, and let $j = 1,
\ldots, J$ denote the $j$th outcome. Let $y_{ij}$ denote the observed
response of participant $i$ on outcome $j$ with marginal distribution
$F_j$, then $y_{ij}$ can be represented as $y_{ij} = F_j^{-1}(u_{ij})$,
where $u_{ij}$ is a uniform $(0,1)$ random variable. An equivalent
representation is $y_{ij} = F_j^{-1} [\Phi(z_{ij}) ]$, where
$\Phi(\cdot)$ denotes the normal CDF and $z_{ij}$ is distributed
standard normal. The unobserved variables $z_{ij}$ are latent responses
underlying each observed response $y_{ij}$.
Assuming that the correlation of $z_{ij}$ with $z_{ij'}$, $1 \leq j <
j' \leq J$,
is specified by the $J \times J$ correlation matrix $\mathbf{C}$, the
Gaussian copula model is
%
%e1 #&#
%e2 #&#
%
\begin{eqnarray}
\mathbf{z}_1, \ldots, \mathbf{z}_n|\mathbf{C} & \sim&
\mbox{i.i.d. } \mathrm{N}(\mathbf{0}, \mathbf{C}),
\\
y_{ij} & =& F_j^{-1} \bigl[\Phi(z_{ij})
\bigr].
\end{eqnarray}
Here, $\mathbf{z}_{i}$ is the $J$-length vector of latent responses
$z_{ij}$ for participant $i$.

In some analyses, the primary focus is on the estimation of the
correlation matrix $\mathbf{C}$ and not the estimation of the marginal
distributions $F_1, \ldots, F_J$. If the latent responses $z_{ij}$ were
known, estimation of $\mathbf{C}$ could proceed using standard methods.
Although the latent responses are unknown, \citet{hoff2007extending}
noted that we do have rank information about the latent responses
through the observed responses because $y_{ij} < y_{ij'}$ implies
$z_{ij} < z_{ij'}$.
If we denote the full set of latent responses by $\mathbf{Z} =
(\mathbf
{z}_1, \ldots, \mathbf{z}_I)^T$ and the full set of observed responses
by $\mathbf{Y} = (\mathbf{y}_1, \ldots, \mathbf{y}_I)^T$, then
$\mathbf
{Z} \in D(\mathbf{Y})$, where
%
%e3 #&#
%
\begin{equation}
\label{eq:rank-set}\quad D(\mathbf{Y})  = \Bigl\{\mathbf{Z} \in\mathbb
{R}^{I \times J} \dvtx\max_{k}\{ z_{kj}\dvtx
y_{kj} < y_{ij}\} < z_{ij} < \min
_{k} \{z_{kj}\dvtx y_{ij} <
y_{kj}\}\ \forall i,j \Bigr\}.
\end{equation}
One can then construct a likelihood for $\mathbf{C}$ that does not
depend on the specification of the marginal distributions $F_1, \ldots,
F_J$ by focusing on the probability of the event, $\mathbf{Z} \in
D(\mathbf{Y})$:
%
%e4 #&#
%
\begin{eqnarray}
\operatorname{Pr} \bigl(\mathbf{Z}\in D(\mathbf{Y})|\mathbf{C},
F_1, \ldots, F_J \bigr) &= &\int_{D(\mathbf{Y})}
p(\mathbf{Z}| \mathbf{C}) \,d\mathbf{Z}
\nonumber
\\[-8pt]
\\[-8pt]
\nonumber
& = &\operatorname{Pr} \bigl(\mathbf{Z}\in D(\mathbf{Y})|\mathbf {C} \bigr).
\end{eqnarray}
Equation (\ref{eq:rank-set}) enables the following decomposition of the
density of $\mathbf{Y}$:
\begin{eqnarray*}
&&p(\mathbf{Y}|\mathbf{C},F_1,\ldots,F_J) \\
&&\qquad= p \bigl(
\mathbf{Y},\mathbf{Z}\in D(\mathbf{Y})|\mathbf{C},F_1,
\ldots,F_J \bigr)
\\
&&\qquad = \operatorname{Pr} \bigl(\mathbf{Z}\in D(\mathbf{Y})|\mathbf{C},F_1,
\ldots,F_J \bigr) \times p \bigl(\mathbf{Y}|\mathbf{Z}\in D(
\mathbf{Y}),\mathbf {C},F_1,\ldots,F_J \bigr)
\\
&&\qquad = \operatorname{Pr} \bigl(\mathbf{Z}\in D(\mathbf{Y})|\mathbf{C} \bigr) \times
p \bigl( \mathbf{Y}|\mathbf{Z}\in D(\mathbf{Y}),\mathbf{C},F_1,
\ldots,F_J \bigr).
\end{eqnarray*}
This decomposition uses the fact that the probability of the event
$\mathbf{Z} \in D(\mathbf{Y})$ does not depend on the marginal
distributions $F_1, \ldots, F_J$ and that the event $\mathbf{Z} \in
D(\mathbf{Y})$ occurs whenever $\mathbf{Y}$ is observed. By using
$\operatorname
{Pr}(\mathbf{Z}\in D(\mathbf{Y})|\mathbf{C})$ as the likelihood
function, the dependence structure of $\mathbf{Y}$ can be estimated
through $\mathbf{C}$ without any knowledge or assumptions about the
marginal distributions. More details on the semiparametric Gaussian
copula model can be found in \citeauthor{hoff2007extending} (\citeyear{hoff2007extending,hoff2009first}).

In the context of latent variable modeling, the main interest is not
just in estimating the correlations among observed variables $\mathbf
{C}$ but in characterizing the interdependencies in multivariate
observed responses through a latent variable model. Latent variable
models, in turn, place constraints on the matrix of correlations among
the observed responses and seek a more parsimonious description of the
dependence structure. Factor analysis is the most common type of latent
variable model with continuous latent variables and continuous outcomes.

To develop a semiparametric approach for factor analysis with mixed
outcomes, assume $Q$ factors, let $\bs{\eta}_i$ be a vector of factor
scores for individual $i$ and $\mathbf{H} = (\bs{\eta}_1, \ldots
, \bs
{\eta}_I )^T$ be the $I \times Q$ factor matrix. Let $\Lambda$
denote the $J \times Q$ matrix of factor loadings. We define our
semiparametric latent variable model as
%
%e5 #&#
%e6 #&#
%e7 #&#
%
\begin{eqnarray}
\bs{\eta}_i & \sim&\mathrm{N}(\mathbf{0}, \mathbf{I}_Q),
\label{eq:spfa-1}
\\
\mathbf{z}_i|\Lambda,\bs{\eta}_i & \sim&\mathrm{N}(
\Lambda\bs{\eta}_i,\mathbf{I}_J), \label{eq:spfa-lat-resp-like}
\\
y_{ij} & =& g_j(z_{ij}).
\label{eq:spfa-4}
\end{eqnarray}
Here, we define $g_j(z_{ij}) = F_j^{-1}(\Phi[z_{ij}/\sqrt{1+\bs
{\lambda}_j^T\bs{\lambda}_j} ] )$, where $\bs{\lambda}_j$
denotes the $j$th row of $\bs{\Lambda}$ and the marginal distribution
$F_j$ remains unspecified. Note that the functions $g_j(\cdot)$ are
nondecreasing and preserve the orderings. The model given by equations~(\ref{eq:spfa-1})--(\ref{eq:spfa-4}) does not rely on the unrestricted
correlation matrix $\mathbf{C}$ as does the Gaussian copula model.
Assuming that a factor analytic model is appropriate for the data, it
constrains the dependencies among the elements of $\mathbf{z}_i$ to be
consistent with the functional form of $\mathbf{I}_J + \bs{\Lambda
}\bs
{\Lambda}^T$. As a result, the proposed semiparametric latent variable
model is a structured case of the semiparametric Gaussian copula model
and can be viewed as a semiparametric form of copula structure
analysis~[\citet{kluppelberg2009copula}].

The general framework of the semiparametric latent variable model given
by equations (\ref{eq:spfa-1})--(\ref{eq:spfa-4}) can be used for any
special cases of factor analysis. In this paper, motivated by the
substantive background information on the SIVD cognitive testing data,
we focus on bifactor models. We define bifactor models as having a
specific structure on the loading matrix, $\bs{\Lambda}$, where each
outcome loads on the primary factor and may load on one or more of the
secondary factors [\citet{reise2007role}]. Most commonly, bifactor
models are applied such that an outcome loads on at most one secondary factor.

%s2.2 #&#
\subsection{Model identification}
The lack of identifiability of factor analysis models that is due to
rotational invariance is well known [\citet{joreskog1969general,dunn1973note,jennrich1978rotational,anderson2003introduction}].
%{\bf Jonathan: Add citations for Joreskog (1969), Dunn (1973, and
%Jennrich (1978). -- see the relabeling paper.}
If we define new factor loadings and new factor scores by $\tilde{\bs
{\Lambda}} = \bs{\Lambda} \mathbf{T}$ and $\tilde{\bs{\eta}}_i =
\mathbf
{T}^{-1}\bs{\eta}_i$, where $\mathbf{T}$ is an orthonormal $Q \times Q$
matrix, then the model
%
%e8 #&#
%
\begin{equation}
\mathbf{z}_i|\bs{\Lambda}, \bs{\eta}_i  \sim \mathrm{N}(
\tilde{\bs{\Lambda}}\tilde{\bs{\eta}}_i, \mathbf{I}_J)
\end{equation}
is indistinguishable from the model in equation~(\ref
{eq:spfa-lat-resp-like}). In the case where the covariance of $\bs
{\eta
}_i$ is not restricted to the identity matrix, any nonsingular $Q
\times Q$ matrix $\mathbf{T}$ results in the same indeterminacy. In
this more general case, we must place $Q^2$ constraints to prevent this
rotational invariance. When we restrict the covariance of $\bs{\eta}_i$
to the identity matrix, this restriction places $\frac12 Q (Q + 1)$
constraints on the model. We are then left with $\frac12 Q(Q-1)$
additional constraints to place on the model. We may satisfy this
requirement by assuming a bifactor structure with $\frac12Q(Q-1)$ zeros
in the matrix of loadings $\bs{\Lambda}$ [\citet{anderson2003introduction}].
%{\bf Jonathan, remove Millsap's citation.}
%By assuming a certain structure for $\bs{\Lambda}$ with zeros as $
%we may satisfy the requirement \citep{anderson2003introduction,
%millsap2001trivial}.
While these restrictions may resolve rotational invariance, the issue
of reflection invariance typically remains [\citet{dunn1973note,jennrich1978rotational}]. Reflection invariance results from the the
fact that the signs of the loadings in any column in $\bs{\Lambda}$ may
be switched.
%and, if the corresponding sign changes are made to the columns in the
%factor matrix, $\mathbf{H}$, then
% & = \mathbf{H} \mathbf{D} \lP\bs{\Lambda} \mathbf{D} \rP^T \\
% & = \tilde{\mathbf{H}} \tilde{\bs{\Lambda}}^T,
Thus, if $\mathbf{D}$ is a diagonal matrix of 1's and $-1$'s
precipitating the sign changes, $\mathbf{H} \bs{\Lambda}^T = \mathbf{H}
\mathbf{D} \mathbf{D} \bs{\Lambda}^T = \tilde{\mathbf{H}} \tilde
{\bs
{\Lambda}}^T$.

% The following paragraph has been heavily revised -- Elena

\citet{geweke1996measuring} proposed an approach that addresses
identifiability of factor models by constraining all upper diagonal
elements in the matrix of factor loadings to zero and requiring all
diagonal elements to be positive. This approach has been used
successfully in Bayesian exploratory factor analysis [\citeauthor{ghosh2008bayesian} (\citeyear{ghosh2008bayesian,ghosh2009default}),
\citet{lopes2004bayesian}], but cannot
be used with bifactor models because the placement of structural zeros
in most cases will be incompatible with fixing all upper diagonal
elements of the matrix of loadings to zero. \citet{congdon2003applied}
and \citet{congdon2006bayesian} suggested the use of a prior that would
place additional constraints on the signs of some of the factor
loadings to resolve the issue of reflection invariance.
However, it has been shown that different choices of parameters for
constraint placement could have a serious impact on model fit in
complex factor models~[\citet{millsap2001trivial}].
%Reflection invariance can be removed by specifying further constraints
%on the signs of selected loadings, however, this approach can be
%problematic in complex factor models where it has been shown that the
%choice of constraint placement can impact model fit
% Milsap's paper
Thus, in our work, we rely on the relabeling algorithm proposed by
\citet
{erosheva2011specification} to resolve reflection invariance. This
algorithm relies on a decision-theoretic approach and resolves the
sign-switching behavior in Bayesian factor analysis in a similar
fashion to the relabeling algorithm introduced to address the
label-switching problem in mixture models [\citet{stephens2000dealing}].
It does not require making preferential choices among variables for
constraint placement.

% Note: this section does not talk about Bayesian estimation
%specifically, so it might be best to not go into details.
%
%While reflection invariance does not typically inhibit the process of
%obtaining maximum likelihood estimates,
%commonly proposed model restrictions to resolve reflection invariance
%may create problems in accurately approximating the posterior
%distributions. To resolve this invariance, we apply the relabeling
%algorithm proposed by \citet{curtis2011specification}. This algorithm
%relies on a decision-theoretic approach to determining the signs and
%resolving this reflection invariance.

In the semiparametric latent variable model, unlike the standard factor
analysis model, specific means and variances are not identifiable. Let
%
%e9 #&#
%
\begin{equation}
\tilde{z}_{ij}  = \mu_j + \sigma_j
z_{ij},
\end{equation}
where $\mu_j$ and $\sigma_j$ are the specific mean and variance for
item $j$. Moreover, if $\tilde{\mathbf{Z}} \in\mathbb{R}^{I \times J}$
denotes the matrix of elements $\tilde{z}_{ij}$ and
%
%e10 #&#
%
\begin{equation}
\tilde{D}(\mathbf{Y})  = \bigl\{\tilde{\mathbf{Z}}\dvtx\max\{
\tilde{z}_{kj}\dvtx y_{kj} < y_{ij}\} <
\tilde{z}_{ij} < \min\{\tilde{z}_{kj}\dvtx y_{ij} <
y_{kj}\} \bigr\},
\end{equation}
then
%
%e11 #&#
%
\begin{equation}
\operatorname{Pr} \bigl(\tilde{\mathbf{Z}} \in\tilde{D}(\mathbf {Y})|\bs{\Lambda
},\mathbf{H},\bs{\mu}, \bs{\Sigma} \bigr)  = \operatorname {Pr} \bigl(\mathbf{Z}
\in D(\mathbf{Y})|\bs{\Lambda}, \mathbf{H} \bigr).
\end{equation}
Thus, shifts in location and scale of the latent responses will not
alter the probability of belonging to the set of feasible latent
response values implied by orderings of the observed responses. As
such, we set the specific means at $\bs{\mu} = 0$ and the specific
variances at $\bs{\Sigma} = \mathbf{I}_J$.

%s3 #&#
\section{Estimation}\label{sec:SPLVM-estimation}

We employ a parameter expansion approach [\citet{liu1998parameter,liu1999parameter}] for Markov chain Monte Carlo (MCMC) sampling,
following the work of \citet{ghosh2009default} on efficient computation
for Bayesian factor analysis.
We found that this method outperforms a Gibbs sampling algorithm with
standard semi-conjugate priors for factor analysis [\citet
{shi1998bayesian,ghosh2009default}] in that it reduces autocorrelation
among the MCMC draws and results in greater effective sample sizes. %
%simulated data example.

%s3.1 #&#
\subsection{Parameter expansion approach}

%priors for factor analysis models often results in poorly behaved
%Gibbs samplers. Specifically, as with hierarchical models, the use of
%proper but diffuse priors often results in slow mixing due to the high
%dependence among parameters. \citet{ghosh2009default} propose a
%parameter expansion approach to remedy this problem.
%accelerate the EM algorithm and \citet{liu1999parameter} applied
%parameter expansion to Gibbs sampling. Among notable applications of
%the parameter expansion technique, \citet{gelman2006prior} applied this
%approach to propose a set of prior distributions for variance
%parameters in hierarchical models as alternatives to the commonly
%adopted inverse gamma prior distribution. Ghosh and Dunson's
%factor analysis models induces $t$ or folded-$t$ prior distributions
%on the factor loadings. Moreover, they demonstrate that the parameter
%expansion approach leads to greater efficiency in sampling than the
%standard Gibbs approach for a number of cases.
%
The central idea behind the parameter expansion approach, using the
terminology of \citet{ghosh2009default}, is to start with a working
model that is an overparameterized version of the initial inferential model.
After proceeding through MCMC sampling, a transformation is used to
relate the draws from the working model to draws from the inferential
model. For our application, the overparameterized model is
%
%e12 #&#
%e13 #&#
%
\begin{eqnarray}
\mathbf{z}_i^\ast& \sim&\mathrm{N} \bigl(\bs{
\Lambda}^\ast\bs{\eta}_i^\ast, \bs{\Sigma}
\bigr),
\\
\bs{\eta}_i^\ast& \sim&\mathrm{N}(\mathbf{0}, \bs{\Psi} ),
\end{eqnarray}
where $\bs{\Sigma}$ and $\bs{\Psi}$ are diagonal matrices that are no
longer restricted to identity matrices. The latent responses $\mathbf
{z}_i^\ast$, the latent variables $\bs{\eta}_i^\ast$ and the loadings
$\bs{\Lambda}^\ast$ are unidentified in this working model. The\vadjust{\goodbreak}
transformations from the working model to the inferential model are
then specified as
%
%e14 #&#
%
\begin{eqnarray}
\nonumber
\bs{\eta}_i & =& \bs{\Psi}^{-1/2}\bs{
\eta}_i^\ast,
\\
\label{eq:spfa-px-transform} \mathbf{z}_i & =& \bs{
\Sigma}^{-1/2}\mathbf{z}_i^\ast,
\\
\nonumber
\bs{\Lambda} & =& \bs{\Sigma}^{-1/2}\bs{\Lambda}^\ast
\bs{\Psi}^{1/2}.
\end{eqnarray}

To sample from the working model, we must specify priors for the
diagonal elements of $\bs{\Psi}$ and $\bs{\Sigma}$ as well as for
$\bs
{\Lambda}^{\ast}$. We specify these priors in terms of the precisions
$\psi_{q}^{-2}$ and $\sigma_{j}^{-2}$. In addition, we denote by $\bs
{\lambda}_j^{\ast\prime}$ the nonzero elements of the $j$th row of
$\bs
{\Lambda}^{\ast}$. The prior on $\bs{\lambda}_j^{\prime}$ is then
induced through the priors on $\psi_{q}^{-2}$, $\sigma_{j}^{-2}$ and
$\bs{\lambda}_j^{\ast\prime}$, rather than being specified directly.
Our priors are
%
%e15 #&#
%
\begin{eqnarray}\label{eq:px-priors}
\nonumber
\psi_{q}^{-2} & \sim&\operatorname{Gamma}(
\phi_{\psi}, \nu_{\psi}),
\\
 \sigma_{j}^{-2} & \sim&
\operatorname{Gamma}(\phi_{\sigma}, \nu_{\sigma}),
\\
\nonumber
\bs{\lambda}_j^{\ast\prime} & \sim&\mathrm{N}(
\mathbf{m}_{\lambda_j^{\ast
\prime}}, \mathbf{S}_{\lambda_j^{\ast\prime}}).
\end{eqnarray}

The structural zeros in the matrix of loadings $\bs{\Lambda}$ are
specified in accordance with our substantive understanding of the
research problem at hand. However, we must have enough zeros so that
the model can be identified since we rely on the placement of these
structural zeros to resolve rotational invariance [\citet
{joreskog1969general,dunn1973note,jennrich1978rotational,loken2005identification}]. Formally, we specify the prior for these
structural zero elements as
%
%e16 #&#
%
\begin{equation}
\lambda_{jq}^{\ast}  \sim\delta_0,
\end{equation}
where $\delta_0$ is a distribution with its point mass concentrated at
0. We estimate loadings with no additional constraints on their signs.
As discussed in Section~\ref{sec:SPLVM-model}, we then deal with
potential multiple modes of the posterior that are due to reflection
invariance by applying the relabeling algorithm proposed by \citet
{erosheva2011specification}.

We now develop the parameter-expanded Gibbs algorithm for sampling
factors $\mathbf{H}$ and loadings $\bs{\Lambda}$. Because the extended
rank likelihood $\operatorname{Pr} (\mathbf{Z}^{\ast}
\in
D(\mathbf{Y})|\bs{\Lambda}^{\ast}, \mathbf{H}^{\ast},  \bs{\Sigma
} )$
involves a complicated integral, any expressions involving it will be
difficult to compute. We avoid having to compute this integral by
drawing from the joint posterior of $(\mathbf{Z}^{\ast}, \mathbf
{H}^{\ast}, \bs{\Lambda}^{\ast}, \bs{\Sigma}, \bs{\Psi})$ via Gibbs
sampling. Given $\mathbf{Z}^{\ast} = \mathbf{z}^{\ast}$ and
$\mathbf
{Z}^{\ast} \in D(\mathbf{Y})$, the full conditional density of $\bs
{\Lambda}^{\ast}$ can be written as
\begin{eqnarray*}
p \bigl(\bs{\Lambda}^{\ast}|\mathbf{H}^{\ast},
\mathbf{Z}^{\ast} = \mathbf{z}^{\ast}, \mathbf{Z}^{\ast} \in
D(\mathbf{Y}), \bs{ \Sigma} \bigr) =p \bigl(\bs{\Lambda}^{\ast}|
\mathbf{H}^{\ast}, \mathbf{Z}^{\ast} = \mathbf{z}^{\ast},
\bs{\Sigma} \bigr)
\end{eqnarray*}
because the current draw values $\mathbf{Z}^{\ast}=\mathbf{z}^{\ast}$
imply $\mathbf{Z}^{\ast}\in D(\mathbf{Y})$. A similar simplification
may be made with the full conditional density of $\mathbf{H}^{\ast}$.
Given $\bs{\Lambda}^{\ast}, \mathbf{H}^{\ast}, \mathbf{Z}^{\ast}
\in
D(\mathbf{Y}), \bs{\Sigma}$ and $\mathbf{Z}_{(-i)(-j)}^{\ast}$, the
full conditional density of $z_{ij}$ is proportional to a normal
density with mean $ (\bs{\lambda}_j^{\ast} )^T\bs{\eta}_i^{\ast
}$ and variance $\sigma_j^2$ that is restricted to the region specified
by $D(\mathbf{Y})$. Our Gibbs sampling procedure for the working model
proceeds according to the following steps:
\begin{longlist}[1.]
\item[1.]\textit{Draw latent responses $\mathbf{Z}^\ast$}. For each $i$
and $j$, sample $z_{ij}^{\ast}$ from a truncated normal distribution
according to
%
%e17 #&#
%
\begin{equation}
z_{ij}^{\ast} \sim\mathrm{TN}_{(z_l^\ast, z_u^\ast)} \bigl( \bigl(\bs {
\lambda }_j^{\ast} \bigr)^T\bs{
\eta}_i^{\ast}, \sigma_{j}^2 \bigr),
\end{equation}
where TN denotes truncated normal and $z_l^\ast, z_u^\ast$ define the
lower and upper truncation points:
%
%e18 #&#
%e19 #&#
%
\begin{eqnarray}
z_{l}^\ast& =& \max_k \bigl
\{z_{kj}^\ast: y_{kj} < y_{ij} \bigr\},
\\
z_{u}^\ast& =& \min_k \bigl
\{z_{kj}^\ast: y_{kj} > y_{ij} \bigr\}.
\end{eqnarray}
\item[2.]\textit{Draw latent variables $\mathbf{H}^\ast$}. For each $i$,
draw directly from the full conditional distribution for $\bs{\eta
}_i^{\ast}$ as follows:
%
%e20 #&#
%
\begin{eqnarray}
\eta_i^{\ast}  &\sim&\mathrm{N} \bigl( \bigl(\bs{
\Psi}^{-1} + \bigl(\bs{\Lambda}^{\ast} \bigr)^T\bs{
\Sigma}^{-1}\bs{ \Lambda}^{\ast} \bigr)^{-1} \bigl(\bs{
\Lambda}^{\ast} \bigr)^T\Sigma^{-1}
\mathbf{z}_i^{\ast},
\nonumber
\\[-8pt]
\\[-8pt]
\nonumber
&&\hspace*{73pt} \bigl( \bs{\Psi}^{-1} + \bigl(
\bs{\Lambda}^{\ast} \bigr)^T\bs{ \Sigma}^{-1}\bs{
\Lambda}^{\ast} \bigr)^{-1} \bigr).
\end{eqnarray}
\item[3.]\textit{Draw loadings $\bs{\Lambda}^{\ast}$}. For each $j$, draw
from the full conditional distribution for the nonzero loadings $\bs
{\lambda}^{\ast\prime}_j$:
%
%e21 #&#
%
\begin{eqnarray}
\bs{\lambda}^{\ast\prime}_j & \sim&\mathrm{N} \bigl( \bigl(
\mathbf{S}_{\lambda
^\prime_j}^{-1} + \sigma_j^{-2}
\bigl( \mathbf{H}_j^{\ast\prime} \bigr)^T\mathbf
{H}_j^{\ast\prime} \bigr)^{-1} \bigl(
\mathbf{S}_{\lambda^\prime
_j}^{-1}m_{\lambda^\prime_j} + \sigma_j^{-2}
\bigl(\mathbf{H}_j^{\ast\prime
} \bigr)^T
\mathbf{z}_j^{\ast} \bigr),
\nonumber
\\[-8pt]
\\[-8pt]
\nonumber
&&\hspace*{131pt}\bigl( \mathbf{S}_{\lambda^\prime_j}^{-1}
+ \sigma_j^{-2} \bigl( \mathbf{H}_j^{\ast\prime}
\bigr)^T\mathbf{H}_j^{\ast
} \prime
\bigr)^{-1} \bigr),
\end{eqnarray}
where $\mathbf{H}_j^{\ast\prime}$ is a matrix comprised of the columns
of $\mathbf{H}^{\ast}$ for which there are nonzero loadings in $\bs
{\lambda}_j$.
\item[4.]\textit{Draw inverse covariance matrix $\bs{\Psi}^{-1}$.} For
each $q$, draw the diagonal element $\psi_q^{-2}$ of $\bs{\Psi}^{-1}$
from the full conditional distribution:
%
%e22 #&#
%
\begin{equation}
\psi_q^{-2}  \sim\operatorname{Gamma} \biggl(
\phi_{\psi} + I/2, \nu_{\psi} + \frac12\sum
_i \eta_{iq}^2 \biggr),
\end{equation}
where $I$ is the number of participants.
\item[5.]\textit{Draw inverse covariance matrix $\bs{\Sigma}^{-1}$.} For
each $j$, draw the diagonal element $\sigma_j^{-2}$ of $\bs{\Sigma
}^{-1}$ from the full conditional distribution:
%
%e23 #&#
%
\begin{equation}
\sigma_j^{-2}  \sim\operatorname{Gamma} \bigl(
\phi_{\sigma} + I/2, \nu_{\sigma} + \tfrac12 (\mathbf{z}_j
- \mathbf{H}\bs{\lambda}_j )^T (\mathbf
{z}_j - \mathbf{H}\bs{\lambda}_j ) \bigr).
\end{equation}
\end{longlist}

After discarding some number of initial draws as burn-in, we transform
the remaining draws using equations~\eqref{eq:spfa-px-transform} as
part of a postprocessing step to obtain
posterior draws from our inferential model. The only remaining steps
are to apply the relabeling algorithm of \citet
{erosheva2011specification}, assess convergence and calculate posterior
summaries for the parameters in the inferential model.
%In the slightly simpler case, where $\bs{\Sigma}$ is restricted to the
%identity matrix but the (diagonal) elements of $\bs{\Psi}$ are
%unrestricted in the working model, we skip the Gibbs sampling step for
%$\bs{\Sigma}$ in the above algorithm.

Our application of parameter expansion to factor analysis models
induces prior distributions that are different from standard
semi-conjugate priors in factor analysis. If the prior covariance
matrix on $\bs{\lambda}_j^\prime$ is diagonal, the prior induced on
$\lambda_{jq}^\prime$ by the parameter expansion is the product of the
normal distribution prior on $\lambda_{jq}^{\ast\prime}$ and the square
root of a ratio of gamma distribution priors on $\sigma_j^{-2}$ and
$\psi_{q}^{-2}$. The ratio of gamma distributed random variables has a
compound gamma distribution which is a form of the generalized beta
prime distribution with the shape parameter fixed to 1. If we integrate
out this ratio, the prior for $\lambda_{jq}$ is a scale mixture of
normals [\citet{west1987scale}] with a compound gamma mixing density. %
%research on shrinkage priors for high-dimensional regression problems
%(see \citealp{armagan2011generalized} and \citealp{polson2010shrink}).
%If the covariance of the latent responses, $\bs{\Sigma}$, is fixed to
%the identity matrix, then the priors induced on the factor loadings
%are $t$-distributions as in \citet{ghosh2009default}.]

The induced prior on the matrix $\bs{\Lambda}$ results in correlations
among elements of the same column and elements of the same row. As
discussed in \citet{ghosh2009default}, prior dependence in the factor
loadings for the $q$th factor will result from the shared parameter
$\psi_q^2$ in their respective prior distributions in the parameter
expanded formulation. Similarly, the shared parameter $\sigma_j^2$ in
the prior distributions for the factor loadings related to the $j$th
outcome in the parameter expanded approach will induce prior dependence
across rows of the factor loadings matrix.

In both the simulation and applied settings considered below,
sensitivity analyses demonstrated that posterior estimates did not
change meaningfully for various hyperparmater values for $\sigma
_j^{-2}$ and $\psi_{q}^{-2}$. For values of $\nu_{\sigma}=1, \phi
_{\sigma}=2, \nu_{\psi}=1/2, \phi_{\psi}=2$, the induced prior on the
nonzero elements of $\bs{\Lambda}$ will have mean, variance and 2.5\%
and 97.5\% quantiles close to that of a standard normal distribution.
In their comparable model, \citet{murray2011bayesian} use a shrinkage
prior, explore its properties and develop a parameter-expanded approach
with optimality properties.%}%Thus, we note that the developed
%parameter expansion approach results in different priors for the
%factor loadings $\bs{\Lambda}$ than the standard priors presented in
%Section~\ref{subsec:stand-gibbs-fa}.

%Thus, while we have approached the use of PX as an alternative means
%of estimation, the application of PX as above results in different
%priors on the factor loadings $\bs{\Lambda}$ and different models
%overall.

%s3.2 #&#
\subsection{Generating replicated data for posterior predictive model checks}
Following \citet{hoff2007extending}, we obtain posterior predictive
distributions that incorporate uncertainty in estimation of the
univariate marginal distributions. Let the superscript $(m)$ denote the
$m$th replicate from the $m$th posterior draw of the parameter. We
generate a new vector of latent responses, $z_{I+1}^{(m)}$, in addition
to $I$ sets drawn as part of the Gibbs sampling algorithm, according to
%
%e24 #&#
%
\begin{equation}
\mathbf{z}_{I+1}^{(m)} \sim\mathrm{N} \bigl(\mathbf{0},
\mathbf{I}_J + \bs{\Lambda}^{(m)} \bigl(\bs{
\Lambda}^{(m)} \bigr)^T \bigr).
\end{equation}
If $z_{(I+1)j}^{(m)}$ falls between two latent responses,
$z_{ij}^{(m)}$ and $z_{i^\prime j}^{(m)}$, that share the same value on
the original data scale (i.e., $y_{ij} = y_{i^\prime j}$), then
$y_{(I+1)j}^{(m)}$ must also take this value as $g_j(\cdot)$ is
monotonic. If $z_{(I+1)j}^{(m)}$ falls between two latent responses,
$z_{ij}^{(m)}$ and $z_{i^\prime j}^{(m)}$, that do not share the same
value on the original data scale, then we select the value, $y_{ij}$ or
$y_{i^\prime j}$, corresponding to the latent response to which
$z_{(I+1)j}^{(m)}$ is closest. In the case of continuous observed
responses, we use linear interpolation to obtain a value for $y_{(I+1)j}^{(m)}$.

%s4 #&#
\section{Hierarchical semiparametric latent variable model}\label
{sec:SPLVM-hierarchical}

To relate covariates of interest to the primary factor, we extend the
proposed model hierarchically. Previously, we assumed that
%
%e25 #&#
%
\begin{equation}
\label{eq:eta-samp-dist} \bs{\eta}_i \sim\mathrm{N}(
\mathbf{m}_{\eta_i}, \bs{\Psi} ),
\end{equation}
where $\mathbf{m}_{\eta_i} = \mathbf{0}, \bs{\Psi} = \mathbf
{I}_Q$. We
now replace the first element of $\mathbf{m}_{\eta_i}$ with a function
of the covariates of interest denoted by the $P$-length vector $\mathbf{x}_i$:
%
%e26 #&#
%
\begin{equation}
\label{eq:eta-samp-dist-mean} \mathbf{m}_{\eta_i}  =
\bigl({x}_i^T \bs{\beta}, 0, \ldots, 0
\bigr)^T. %\begin{pmatrix}
% \\
%0
\end{equation}
When we employ a parameter expansion approach for estimation,
%
%e27 #&#
%e28 #&#
%
\begin{eqnarray}
\label{eq:px-hier-dists} \bs{\eta}_i^{\ast} & \sim&\mathrm{N}
(\mathbf{m}_{\eta_i}, \bs{\Psi} ),
\\
\mathbf{m}_{\eta_i} & =& \bigl({x}_i^T\bs{
\beta}^{\ast}, 0, \ldots, 0 \bigr)^T,
\end{eqnarray}
where the diagonal elements of $\bs{\Psi}$ are no longer restricted
during MCMC.
For $\bs{\beta}^{\ast}$, we specify the semi-conjugate prior:
%
%e29 #&#
%
\begin{equation}
\label{eq:beta-samp-dist} \bs{\beta}^{\ast}  \sim\mathrm{N} (
\mathbf{m}_{\beta}, \mathbf{S}_{\beta} ).
\end{equation}

Moreover, to further facilitate efficient computation, we add an
additional working parameter, $\alpha$, as suggested by \citet
{ghosh2009default}, so that
%
%e30 #&#
%
\begin{equation}
\mathbf{m}_{\eta_i}  = \bigl(\alpha+ {x}_i^T\bs{
\beta}^{\ast}, 0, \ldots, 0 \bigr)^T. %\eta_{i1}^{\ast} & \sim
\end{equation}
Relaxing the restriction on the mean of the latent variable promotes
better mixing of the regression coefficients. For $\alpha$, we use the
semi-conjugate prior:
%
%e31 #&#
%
\begin{equation}
\alpha \sim\mathrm{N} \bigl(m_{\alpha}, s_{\alpha}^2
\bigr).
\end{equation}

To estimate the hierarchical model [equations~\eqref{eq:eta-samp-dist},
\eqref{eq:eta-samp-dist-mean}], we modify the steps for drawing $\bs
{\eta}_i^{\ast}$ and $\bs{\Psi}$ in the sampling algorithm from
Section~\ref{sec:SPLVM-estimation} to account for the inclusion of
covariates and the additional working parameter, $\alpha$, in $\mathbf
{m}_{\eta_i}$. We sample $\bs{\beta}^{\ast}$ and $\alpha$
according to
their full conditionals:
%
%e32 #&#
%e33 #&#
%
\begin{eqnarray}
\bs{\beta}^{\ast} & \sim&\mathrm{N} \bigl( \bigl(\psi_1^{-2}
\mathbf{X}^T\mathbf{X} + \mathbf{S}_{\beta}^{-1}
\bigr)^{-1} \bigl(\psi_1^{-2}\mathbf{X}^T
\bigl( \bs{\eta}_{q=1}^{\ast} - \mathbf{1}_I\alpha
\bigr)+ \mathbf{S}_{\beta
}^{-1}\mathbf{m}_{\beta} \bigr),
\nonumber
\\[-8pt]
\\[-8pt]
\nonumber
&&\hspace*{162pt}
\bigl( \psi_1^{-2}\mathbf{X}^T\mathbf{X} +
\mathbf{S}_{\beta}^{-1} \bigr)^{-1} \bigr),
\\
\alpha& \sim&\mathrm{N} \bigl( \bigl(\psi_1^{-2}I +
s_{\alpha}^{-2} \bigr)^{-1} \bigl(\psi_1^{-2}
\mathbf{1}_I^T \bigl(\bs{\eta}_{q=1}^{\ast}
- \mathbf{X}\bs{\beta}^{\ast} \bigr)+ s_{\alpha}^{-2}m_{\alpha}
\bigr),
\nonumber
\\[-8pt]
\\[-8pt]
\nonumber
&&\hspace*{161pt} \bigl(\psi_1^{-2}I + s_{\alpha}^{-2}
\bigr)^{-1} \bigr),
\end{eqnarray}
where $\mathbf{X}$ is an $I \times P$ matrix of covariates and $\bs
{\eta
}_{q=1}^{\ast}$ is a vector of the primary factor scores, the first
column of $\mathbf{H}^{\ast}$.
In the postprocessing stage for the parameter expansion approach, we
make the transformations:
%
%e34 #&#
%e35 #&#
%
\begin{eqnarray}
\bs{\eta}_{i} & =& \bs{\Psi}^{-1/2} \bigl(\bs{
\eta}_i^\ast- \bs{\alpha} \bigr),
\\
\bs{\beta} & = &\bs{\beta}^\ast\psi_1^{-1},
\end{eqnarray}
where $\bs{\alpha} = (\alpha, 0, \ldots, 0)^T$. %In the parameter
%expansion approach, given diagonal $S_{\beta}$, the induced prior on $
%density. This normal-gamma prior was proposed by
%that generalizes the Bayesian Lasso \citep{park2008bayesian}.

%In Section~\ref{sec:SPLVM-estimation}, we presented two alternatives
%for generating replicated data using the posterior predictive
%distribution. The first method relied on the fact that we may
%integrate out the value of $\bs{\eta}_i$ so that the distribution of $
%such as $\bs{\eta}_i$. However, in the hierarchical formulation, this
%is no longer the case. Even if we integrate out $\bs{\eta}_i$, the
%distribution of $\mathbf{z}_i$ will depend on $\mathbf{x}_i$. As a
%result, we rely on the second method of sampling from the posterior
%predictive distribution to perform posterior predictive model
%checking. We first resample $I$ vectors of covariates from $
%responses $\mathbf{Z}^{(\mathrm{rep},m)}$ using the posterior draws of
%$
%the original scale by comparing the predicted latent responses to
%those generated by the Gibbs sampler as described above.

%s5 #&#
\section{Simulation data example and application to SIVD data}\label
{sec:SIVD-app}

%s5.1 #&#
\subsection{Simulated data example}
To test the semiparametric latent variable model, we examined the
model's ability to recover data generating parameters using simulated
data for $I=500$ individuals on $J=15$ outcomes. We applied three
estimation approaches: the parameter expansion approach from
Section~\ref{sec:SPLVM-estimation}, a variation of the parameter
expansion approach where only the diagonal elements of $\bs{\Psi}$ are
unrestricted during estimation, and a more standard Gibbs sampling
approach [\citet{shi1998bayesian,ghosh2009default}].
The data generating process assumed a bifactor form with two secondary
factors in addition to the primary factor.
All outcomes loaded on the general factor; outcomes 1, 13, 14 and 15
loaded on one secondary factor; and outcomes 3, 4, 6 and 8 loaded on
the other secondary factor. For each individual, we simulated $\bs
{\eta
}_i \sim \mathrm{N}(\mathbf{0}, \mathbf{I}_Q)$. We subsequently generated a
matrix of latent responses, $\mathbf{Z}$, with mean $\mathbf{H}^T\bs
{\Lambda}$. Finally, we randomly drew cutoffs for each outcome in order
to produce discretized ``observed'' responses from the continuous
latent responses so that the number of unique values for each outcome
ranged from 2 (outcome 1) to 30 (outcome 9).

To fit the model to the simulated data, we employed each estimation
approach to generate 50,000 MCMC draws, the first 10,000 of which we
discarded as burn-in. In addition, we thinned the posterior samples,
keeping only every 10th draw. All estimation approaches did a good job
of recovering the data generating values, but the parameter expansion
estimation approach described in Section~\ref{sec:SPLVM-estimation}
displayed better mixing, less autocorrelation and larger effective
sample sizes, sometimes by a factor of ten or more compared to a
standard Gibbs sampling approach. We note that because the target
distributions are not the same for the three approaches, measures such
as effective sample size are not directly comparable. However, as in
\citet{ghosh2009default}, we use these measures as indicators of the
quality of mixing using the different approaches. Finally, as discussed
in Section~\ref{sec:SPLVM-estimation}, different choices for
hyperparameter values did not result in meaningful differences in the
posterior estimates.
%We provide a complete description of the simulated data example in

%s5.2 #&#
\subsection{Application to SIVD data}

Participants were recruited to fill six broad groups in the Subcortical
Ischemic Vascular Dementia (SIVD) study [\citet{chui2007subcortical}],
comprised by three levels of cognitive functioning and two levels
(absence vs. presence) of subcortical lacunes. Lacunes are small areas
of dead brain tissue caused by blocked or restricted blood supply. The
three levels of cognitive functioning groups were normal, mildly
impaired and demented as determined by the Clinical Dementia Rating
total score, a numerical rating that is based on medical history and
clinical examination as well as other forms of assessment [\citeauthor{morris1993clinical}
(\citeyear{morris1993clinical,morris1997clinical})]. Among the data collected by
SIVD are neuropsychological test results and standardized magnetic
resonance imaging (MRI) scans of the participants' brains [\citet
{mungas2005longitudinal}]. A computerized segmentation algorithm
classified pixels from the MRI scans into different components,
including white matter hyperintensities [\citet{cardenas2001reliability}].

We are interested in relating the individuals' level of executive
functioning to the white matter hyperintensity volume located in the
frontal lobe at individuals' first visit. White matter hyperintensities
are areas of increased signal intensity that are commonly associated
with older age. Among the outcomes available at first visit, we
identified 21 indicators of executive functioning in the SIVD
neuropsychological battery. These items included Digit Span, Visual
Span, Verbal Fluency, Stroop Test and
Mattis Dementia Rating Scale (MDRS) test items.
We excluded MDRS outcomes M and N, as everyone except one participant
received full credit on these outcomes. As a result, we used 19 of the
21 executive functioning outcomes in our analysis.

Table~\ref{tab:sivditems} displays basic information for the 19
outcomes as well as some summary statistics observed in the data for $I
= 341$ participants. For this analysis, we considered only participants
with a complete set of responses to the 19 executive functioning
outcomes as well as a concurrent set of brain MRI measurements. We
defined concurrent as within six months (before or after) of the
neuropsychological testing date.
As one can see from the summary statistics, the outcomes vary greatly
in their number of categories as well as in their difficulty. For many
of the binary outcomes as well as the MDRS outcomes E and V, the mean
and median scores are very close to the largest possible score.

%t1 #&#
%
\begin{table}
\caption{Summary statistics for $I=341$ responses to 19 SIVD executive
functioning outcomes as well as outcome type assignment. ``RC Count''
denotes a right-censored count outcome}
\label{tab:sivditems}
\begin{tabular*}{\textwidth}{@{\extracolsep{\fill}}lcd{2.2}d{2.0}c@{}}
\hline
& \textbf{Range} & \multicolumn{1}{c}{\textbf{Mean}} & \multicolumn{1}{c}{\textbf{Median}} & \multicolumn{1}{c@{}}{\textbf{Outcome type}} \\
\hline
Digit Span Forward & 3--12 & 7.69 & 8 & Count \\
Digit Span Backward & 1--12 & 5.97 & 6 & Count \\
Visual Span Forward & 0--13 & 7.15 & 7 & Count \\
Visual Span Backward & 0--12 & 6.18 & 6 & Count \\
Verbal Fluency Letter F & 1--26 & 11.8 & 12 & Count \\
Verbal Fluency Letter A & 0--40 & 10.2 & 10 & Count \\
Verbal Fluency Letter S & 0--50 & 12.4 & 12 & Count \\
MDRS E & 2--20 & 16.64 & 19 & RC Count \\
MDRS G & 0--1 & 0.96 & 1 & Binary \\
MDRS H & 0--1 & 0.98 & 1 & Binary \\
MDRS I & 0--1 & 0.95 & 1 & Binary \\
MDRS J & 0--1 & 0.97 & 1 & Binary \\
MDRS K & 0--1 & 0.98 & 1 & Binary \\
MDRS L & 0--1 & 0.79 & 1 & Binary \\
MDRS O & 0--1 & 0.94 & 1 & Binary \\
MDRS V & 9--16 & 14.9 & 16 & RC Count \\
MDRS W & 0--8 & 6.44 & 7 & Ordered Cat. \\
MDRS X & 0--3 & 2.66 & 3 & Ordered Cat. \\
MDRS Y & 0--3 & 2.93 & 3 & Ordered Cat. \\
\hline
\end{tabular*}
\end{table}

\begin{figure}

\includegraphics{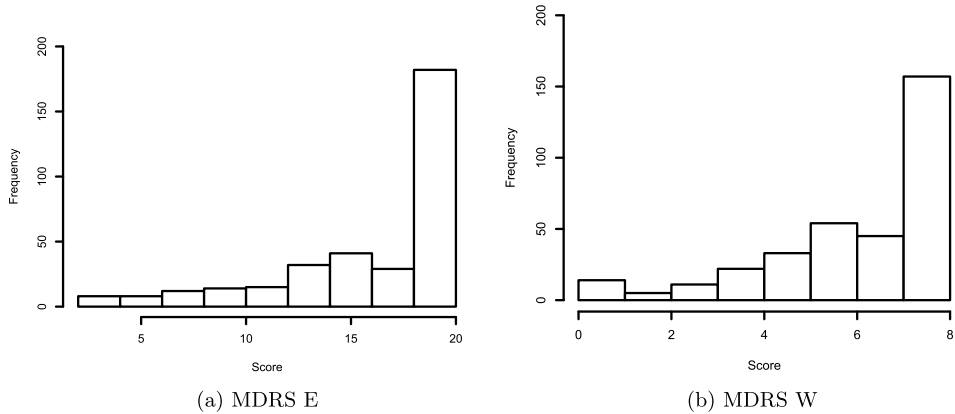}

\caption{Histograms of scores for MDRS E and W items.}
\label{fig:hists}\vspace*{-3pt}
\end{figure}

To illustrate the challenges of modeling cognitive outcomes from the
SIVD study parametrically, we describe two items in more detail. For
MDRS outcome~E, participants are given one minute and are asked to name
as many items found in supermarkets as they can.
The participant's score is the number of valid items named, censored at
20. A histogram of observed scores for this outcome in Figure~\ref{fig:hists}(a) shows some evidence of a ceiling effect for this item.
Similarly, Figure~\ref{fig:hists}(b) depicts a histogram of observed
scores for MDRS outcome W that asks a participant to compare words and
identify similarities. Although the description in this case does not
suggest right-censoring, there is also some evidence of a ceiling
effect in the histogram. We might treat MDRS outcome W as
right-censored rather than an ordered categorical outcome in a
parametric approach. These are just two examples that illustrate
ambiguities in specifying appropriate parametric distributions for each
cognitive outcome in the SIVD study. To bypass this specification, yet
still model the interdependencies among test items, we used the
hierarchical semiparametric latent variable model.

%f1 #&#
%

%[width=0.45\textwidth]{hist-mdrse.pdf}} \subfloat[][MDRS W]{\label
%{fig:hist-mdrsw}\includegraphics[width=0.45\textwidth]{hist-mdrsw.pdf}}
%

We are interested in modeling the relationship between the primary
factor and the volume of white matter hyperintensities located in the
frontal lobe of the brain. Controlling for other covariates, we
specified the mean of the primary factor as
%
%e36 #&#
%
\begin{equation}
\E[\eta_{i1}] = \beta_1\mathrm{Sex}_i +
\beta_2\mathrm{Educ}_i + \beta_3
\mathrm{Age}_i + \beta_4\mathrm{Vol}_i +
\beta_5\mathrm{WMH}_i,
\end{equation}
where Sex is the participant's sex (Female${}={}$1, Male${}={}$0), Educ is the
number of years of education, Vol is the total brain volume of the
participant, and WMH is the frontal white matter hyperintensity volume.
We used standardized versions of the continuous predictor variables.
Table~\ref{tab:covar-sum} displays some summary statistics for these
covariates by different levels of frontal white matter hyperintensity
volume.%\footnote{We created these groups for presentation purposes
%solely. We used continuous measurement values for frontal white matter
%hyperintensity volume in our models.}

%t2 #&#
%
\begin{table}[b]\vspace*{-3pt}
\caption{Mean and SD for covariates by level of frontal WMH. The range
of frontal WMH measurements was partitioned to obtain three similarly
sized groups}
\label{tab:covar-sum}
\begin{tabular*}{\textwidth}{@{\extracolsep{\fill}}lccc@{}}
\hline
& \multicolumn{3}{c@{}}{\textbf{Frontal WMH (cc)}} \\[-6pt]
& \multicolumn{3}{c@{}}{\hrulefill} \\
& \multicolumn{1}{c}{\textbf{0--5}} & \multicolumn{1}{c}{$\bolds{>}$\textbf{5--11}} & \multicolumn{1}{c@{}}{$\bolds{>}$\textbf{11}} \\
\hline
No. participants & 113 & 112 & 116 \\[3pt]
Age (Yrs) & 68.49 (8.65) & 74.82 (6.72) & 79.44 (6.23) \\
Education (Yrs) & 15.36 (2.95) & 15.12 (3.01) & 14.32 (3.12) \\
Total brain volume (cc) & 1196.2 (115.81) & 1231.65 (114.78) & 1218.34
(138.13) \\
\hline
\end{tabular*}
\end{table}

\textit{One-factor semiparametric model}.
We started our analysis by examining the $Q=1$ model with a single
latent factor explaining interdependencies among the test items.
%We start our analysis by examining the $Q=1$ model with a single
%latent factor, the cognitive ability, explaining interdependencies
%among the 19 outcomes.
To estimate the model, we utilized the parameter-expanded Gibbs
sampling algorithm. Even though we found this approach to be more
efficient than the standard Gibbs sampler, we still observed high
autocorrelation within the chains for factor loadings.
%Even with the improved performance over the Gibbs sampling algorithm
%for the standard model, there is high autocorrelation within the
%factor loadings.
We drew 50,000 MCMC samples and discarded the first half as burn-in. We
used trace plots and the Geweke [\citet{geweke1992evaluating}] and
Raftery--Lewis [\citet{raftery1995number}] diagnostic tests to assess
convergence.
%
% Jonathan: Comment on convergence. List methods you used to assess
%convergence and comment on each.
%
%%
%factor, $Q=1$, and bifactor, $Q=4$, models.}
%%
%%
%& \multicolumn{}
%Coefficient & Mean & Median & 95\% CI \\
%Sex & 0.234 & 0.233 & (-0.061, 0.516) \\
%Education & 0.354 & 0.354 & (0.232, 0.479) \\
%Age & -0.126 & -0.126 & (-0.246, 0.004) \\
%Total Brain Vol. & 0.069 & 0.069 & (-0.080, 0.215) \\
%Frontal WMH Vol. & -0.330 & -0.328 & (-0.466, -0.205) \\
%%
%%
%%

Table~\ref{tab:spfa-reg-coef} displays posterior summaries for the
regression coefficients, $\bs{\beta}$.
We observed a negative relationship between the primary factor and
frontal white matter hyperintensity volume.
%Based on our semiparametric latent variable model, we expect a 1SD
%increase in frontal white matter hyperintensity volume to be
%associated with a 0.330SD decrease in the primary factor.
The accompanying 95\% posterior credible interval ($-0.466, -0.205$) did
not contain zero, suggesting a negative association between frontal
white matter hyperintensity volume and the primary factor.

\begin{figure}

\includegraphics{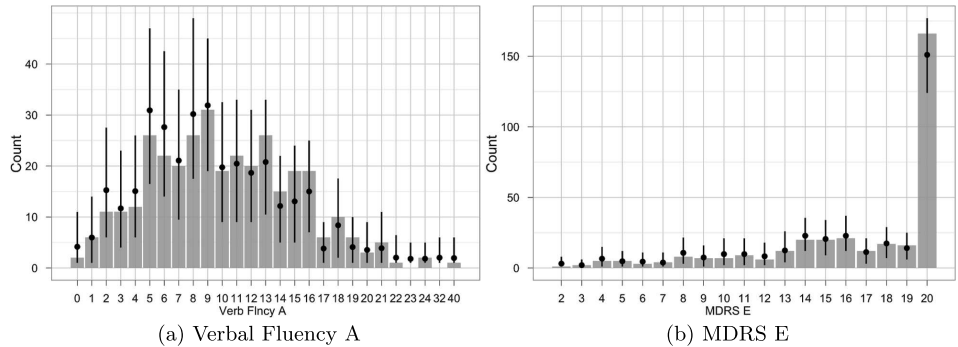}

\caption{Histograms of the observed scores for the Verbal Fluency $A$ and
MDRS $E$. The black points indicate the mean count across replicated data
sets for each score. The black vertical segment indicates the interval
from the 2.5\% to 97.5\% quantiles across replicated data sets.}
\label{fig:spfa-sivd-hist-marg}
\end{figure}

We evaluated model fit using posterior predictive model checks. We
began by examining the fit of the marginal distributions.\vadjust{\goodbreak} Figure~\ref{fig:spfa-sivd-hist-marg} displays the histograms of observed responses
for Verbal Fluency outcome $A$ and MDRS outcome $E$ along with posterior
predictive summaries. In each case, the model appeared to do a
satisfactory job of approximating the data. We found similarly good
approximations of the marginal distributions in the observed data for
the other outcomes as well.
%
%f2 #&#
%

%]{SIVD-Q1-pp-hist-outcome-13.eps}}
%]{SIVD-Q1-pp-hist-outcome-15.eps}}
%

%We now evaluate the model's ability to approximate the dependence
%structure among the observed item responses.

\begin{figure}

\includegraphics{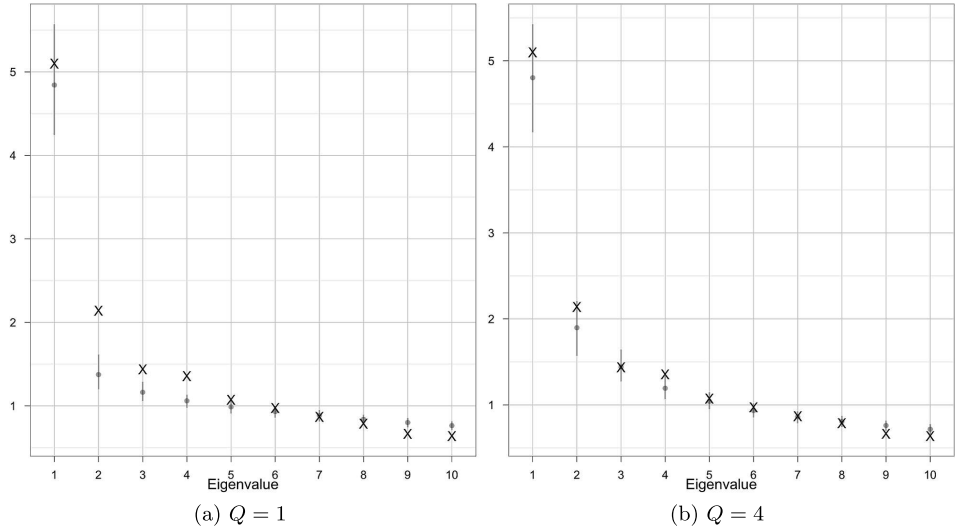}

\caption{Eigenvalue plots for the $Q=1$ and $Q=4$ models. The mean
posterior prediction (grey point) and 95\% posterior prediction
intervals (grey line segment) of the top ten eigenvalues calculated
using replicated data from the single factor ($Q=1$) and bifactor
($Q=4$) models. Eigenvalues computed from the observed data are denoted
by a black~``$X$.''}
\label{fig:eigs}
\end{figure}

We assessed the model's ability to replicate the observed dependence
structure in the data at a global level by examining the eigenvalues of
the observed rank correlation matrix [Figure~\ref{fig:eigs}(a)].
%provides us with information on the model's ability to represent the
%dependence structure.
The eigenvalues of correlation matrices form the basis of heuristic
tests in factor analysis such as the latent root criterion [\citet
{guttman1954some}] or the scree test [\citet{cattell1966scree}] that
determine the number of factors to include in the model. The first
eigenvalue was well approximated by the model but the subsequent
eigenvalues indicated model misfit, suggesting that additional factors
may be necessary to more accurately represent the dependence structure
in the data.

%f3 #&#
%

%
%%
%prediction intervals (grey line segment) of the top ten eigenvalues
%calculated using replicated data from the single factor ($Q=1$) model.
%Eigenvalues computed from the observed data are denoted by a black
%``X''.}
%%

\begin{figure}

\includegraphics{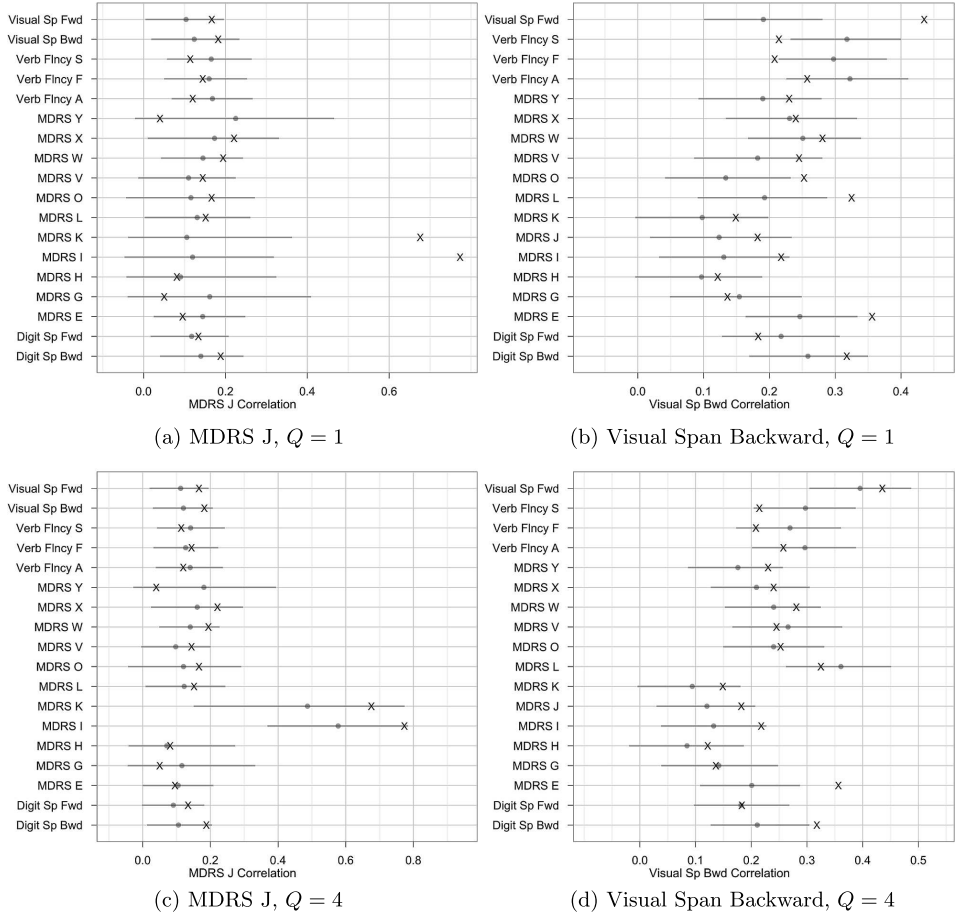}

\caption{Pairwise correlation plots for the single factor ($Q=1$) and
bifactor models ($Q=4$). Each pairwise correlation plot depicts the
mean posterior prediction (grey point) and 95\% posterior prediction
intervals (grey line segment) for Kendall's $\tau$ values calculated
using replicated data. Kendall's $\tau$ values computed from the
observed data are denoted by a black ``X.''}
\label{fig:spfa-sivd-pair-corr}
\end{figure}

We reviewed the pairwise rank correlations to better understand the
shortcomings of the single factor model and direct the next steps in
our model building process. Figures~\ref{fig:spfa-sivd-pair-corr}(a)
and \ref{fig:spfa-sivd-pair-corr}(b) display the pairwise
correlation plots for the MDRS J and Visual Span Backward outcomes for
the single factor model. In both cases, the model fit the majority of
the pairwise correlations well. However, in each case, there were a few
outcomes with poorly fitted correlations. For MDRS J, the model did not
appear to fully capture the correlation with the conceptually-related
MDRS I and K; all three of these outcomes ask participants to repeat
alternating movements of some type. Likewise, for Visual Span Backward,
the correlation with Visual Span Forward was not accurately
approximated by the single factor model. In addition, the correlations
between Visual Span Backward and the MDRS outcomes L and O were not
well approximated. MDRS outcomes L and O involve copying drawings and,
in this sense, also incorporate a visual component that may be the
source of the residual correlation between the outcomes.
The observation that the lack of fit was present among conceptually
related outcomes (e.g., outcomes that are parts of a subtest or a
subscale) is consistent with the notion that possible secondary factors
may impact item correlations in addition to the general executive
functioning factor. Thus, our next step was to consider the class of
bifactor models.

%Given that the lack of fit is observed among outcomes that are related
%conceptually (e.g., that are parts of a subtest or a subscale), our
%next step was to consider a bifactor model.

%%
%]{SIVD-Q1-pp-oc-corrs-4.eps}}
%]{SIVD-Q1-pp-oc-corrs-11.eps}}
%prediction (grey point) and 95\% posterior prediction intervals (grey
%line segment) for Kendall's $\tau$ values calculated using replicated
%data. Kendall's $\tau$ values computed from the observed data are
%denoted by a black ``X''.}
%%
%
%f4 #&#
%

%{fig:splvm-sivd-pair-corr-4-q1}\includegraphics[width=0.45\textwidth
%]{SIVD-Q1-pp-oc-corrs-4.eps}}
%{fig:splvm-sivd-pair-corr-11-q1}\includegraphics[width=0.45\textwidth
%]{SIVD-Q1-pp-oc-corrs-11.eps}}\\
%{fig:splvm-sivd-pair-corr-4-q4}\includegraphics[width=0.45\textwidth
%]{SIVD-Q4-pp-oc-corrs-4.eps}}
%{fig:splvm-sivd-pair-corr-11-q4}\includegraphics[width=0.45\textwidth
%]{SIVD-Q4-pp-oc-corrs-11.eps}}
%

\textit{Bifactor semiparametric model}.
To choose a secondary factors structure in a bifactor model, we applied
an iterative process. During one iteration, we examined all pairwise
correlations for the lack of fit, specified secondary factors to
account for residual correlation, refit the model and checked the fit
of this new model.
%Examining the pairwise correlations for the other items in this
%manner, we identified possible secondary factors to account for
%residual correlation. We applied an iterative process where we
%specified an additional secondary factor, refit the model and then
%checked the fit of this new model.
Ultimately, we specified a bifactor model with one general cognitive
ability factor and 3 secondary factors (for a total of $Q = 4$) as
listed in Table~\ref{tab:SIVD-fact-st}.
%The factor structure is listed in Table~\ref{tab:SIVD-fact-st} where $
%
%%
%outcomes. $\ast$ indicates a nonzero factor loading to be estimated.}
%%
%%
%& \multicolumn{4}{|c}{Factor} \\
%& 1 & 2 & 3 & 4 \\
%MDRS G & $\ast$ & 0 & 0 & 0 \\
%MDRS H & $\ast$ & 0 & 0 & 0 \\
%MDRS I & $\ast$ & $\ast$ & 0 & 0 \\
%MDRS J & $\ast$ & $\ast$ & 0 & 0 \\
%MDRS K & $\ast$ & $\ast$ & 0 & 0 \\
%MDRS L & $\ast$ & 0 & $\ast$ & 0 \\
%MDRS O & $\ast$ & 0 & $\ast$ & 0 \\
%Digit Sp Fwd & $\ast$ & 0 & 0 & 0 \\
%Digit Sp Bwd & $\ast$ & 0 & 0 & 0 \\
%Visual Sp Fwd & $\ast$ & 0 & $\ast$ & 0 \\
%Visual Sp Bwd & $\ast$ & 0 & $\ast$ & 0 \\
%Verb Flncy F & $\ast$ & 0 & 0 & 0 \\
%Verb Flncy A & $\ast$ & 0 & 0 & 0 \\
%Verb Flncy S & $\ast$ & 0 & 0 & 0 \\
%MDRS E & $\ast$ & 0 & 0 & 0 \\
%MDRS V & $\ast$ & 0 & $\ast$ & 0 \\
%MDRS W & $\ast$ & 0 & 0 & $\ast$ \\
%MDRS X & $\ast$ & 0 & 0 & $\ast$ \\
%MDRS Y & $\ast$ & 0 & 0 & $\ast$ \\
%%
%%
%%
%
It is important to note that, although we identified these secondary
factors using the posterior predictive model checks, they nevertheless
have substantive interpretations as they link conceptually related
outcomes. The second factor loads on MDRS outcomes I, J and K, test
items that all involve repetition of alternating movements. The third
factor loads on the Visual Span outcomes and MDRS outcomes L, O and V.
These test items all include visual or drawing components. The fourth
factor links three MDRS outcomes that ask participants to identify
similarities and dissimilarities.

%
%t3 #&#
%
\begin{table}
\tabcolsep=0pt
\caption{Proposed factor structure for SIVD executive functioning
outcomes. $\ast$ indicates a nonzero factor loading to be estimated}\label{tab:SIVD-fact-st}
\begin{tabular*}{\textwidth}{@{\extracolsep{\fill}}lccccccccccccccccccc@{}}
\hline
& \multicolumn{19}{c}{\textbf{Outcomes}} \\[-3pt]
& \multicolumn{19}{c}{\hrulefill}    \\
\textbf{Factor} & \rotatebox{90}{\textbf{MDRS G}} & \rotatebox{90}{\textbf{MDRS H}} & \rotatebox
{90}{\textbf{MDRS I}} & \rotatebox{90}{\textbf{MDRS J}} & \rotatebox{90}{\textbf{MDRS K}} &
\rotatebox{90}{\textbf{MDRS L}} & \rotatebox{90}{\textbf{MDRS O}} & \rotatebox{90}{\textbf{Digit
Sp Fwd}} & \rotatebox{90}{\textbf{Digit Sp Bwd}} & \rotatebox{90}{\textbf{Visual Sp Fwd}}
& \rotatebox{90}{\textbf{Visual Sp Bwd}} & \rotatebox{90}{\textbf{Verb Flncy F}} &
\rotatebox{90}{\textbf{Verb Flncy A}} & \rotatebox{90}{\textbf{Verb Flncy S}} &
\rotatebox
{90}{\textbf{MDRS E}} & \rotatebox{90}{\textbf{MDRS V}} & \rotatebox{90}{\textbf{MDRS W}} &
\rotatebox{90}{\textbf{MDRS X}} & \rotatebox{90}{\textbf{MDRS Y}} \\
\hline
1 & $\ast$ & $\ast$ & $\ast$ & $\ast$ & $\ast$ & $\ast$ & $\ast$
& $\ast
$ & $\ast$ & $\ast$ & $\ast$ & $\ast$ & $\ast$ & $\ast$ & $\ast$
& $\ast
$ & $\ast$ & $\ast$ & $\ast$ \\
2 & 0 & 0 & $\ast$ & $\ast$ & $\ast$ & 0 & 0 & 0 & 0 & 0 & 0 & 0 & 0 &
0 & 0 & 0 & 0 & 0 & 0 \\
3 & 0 & 0 & 0 & 0 & 0 & $\ast$ & $\ast$ & 0 & 0 & $\ast$ & $\ast$ &
0 &
0 & 0 & 0 & $\ast$ & 0 & 0 & 0 \\
4 & 0 & 0 & 0 & 0 & 0 & 0 & 0 & 0 & 0 & 0 & 0 & 0 & 0 & 0 & 0 & 0 &
$\ast$ & $\ast$ & $\ast$ \\
\hline
\end{tabular*}
\end{table}

%
%t4 #&#
%
\begin{table}[b]
\tabcolsep=0pt
\caption{Posterior summaries for regression coefficients for single
factor, $Q=1$, and bifactor, $Q=4$, models}
\label{tab:spfa-reg-coef}
\begin{tabular*}{\textwidth}{@{\extracolsep{\fill}}ld{2.3}d{2.3}cd{2.3}d{2.3}c@{}}
\hline
& \multicolumn{3}{c}{$\bolds{Q=1}$} & \multicolumn{3}{c@{}}{$\bolds{Q=4}$}
\\[-6pt]
& \multicolumn{3}{c}{\hrulefill} & \multicolumn{3}{c@{}}{\hrulefill} \\
\multicolumn{1}{@{}l}{\textbf{Coefficient}} & \multicolumn{1}{c}{\textbf{Mean}} & \multicolumn{1}{c}{\textbf{Median}} &
\multicolumn{1}{c}{\textbf{95\% CI}} & \multicolumn{1}{c}{\textbf{Mean}}& \multicolumn{1}{c}{\textbf{Median}} & \multicolumn{1}{c@{}}{\textbf{95\% CI}}\\
\hline
Sex & 0.234 & 0.233 & $(-0.061, 0.516)$ & 0.155 & 0.152 & $(-0.134, 0.440)$
\\
Education & 0.354 & 0.354 & (0.232, 0.479) & 0.325 & 0.325 & (0.206,
0.446) \\
Age & -0.126 & -0.126 & $(-0.246, 0.004)$ & -0.078 & -0.078 & $(-0.201,
0.044)$ \\
Total brain Vol. & 0.069 & 0.069 & $(-0.080, 0.215)$ & 0.046 & 0.045 &
$(-0.096, 0.194)$ \\
Frontal WMH Vol. & -0.330 & -0.328 & $(-0.466, -0.205)$ & -0.335 & -0.336
& $(-0.464, -0.208)$ \\
\hline
\end{tabular*}
\end{table}

For the semiparametric bifactor model with $Q=4$, we drew 500,000 MCMC
samples and discarded the first 50,000 as burn-in. We kept every 50th
draw, leaving us with 9000 posterior draws. As with the single factor
model, we checked convergence using trace plots and the Geweke [\citet
{geweke1992evaluating}] and Raftery--Lewis [\citet{raftery1995number}]
diagnostic tests. Convergence was satisfactory but, compared to the
single factor model, the mixing was considerably slower for a few of
the secondary factors that exhibited high levels of autocorrelation. We
should also note that the speed of convergence was influenced by the
choice of hyperparameters for $\bs{\Sigma}$ and $\bs{\Psi}$ in the
parameter expanded model.

The bifactor model represented the dependence structure of the observed
responses better. Figure~\ref{fig:eigs}(b) shows that
the bifactor model provided a good fit to the observed eigenvalues well
beyond the first eigenvalue. As can be seen in Figures~\ref{fig:spfa-sivd-pair-corr}(c) and \ref
{fig:spfa-sivd-pair-corr}(d), the bifactor model did a better job
of replicating the pairwise rank correlations compared to the single
factor model. In Figure~\ref{fig:spfa-sivd-pair-corr}(c), one may
also notice the larger posterior credible intervals for the pairwise
rank correlations of MDRS I with MDRS J and MDRS K. Referring back to
Table~\ref{tab:sivditems}, we see that almost all participants answered
these items correctly. As a result, there is less information to
estimate the secondary factor that accounts for the residual
correlation among these three items and, in the wide credible
intervals, we see the subsequent imprecision.

%%
%and 95\% posterior prediction intervals (grey line segment) of the
%eigenvalues calculated using replicated data. Eigenvalues computed from
%the observed data are denoted by a black ``X''.}
%%
%]{SIVD-Q4-pp-oc-corrs-4.eps}}
%]{SIVD-Q4-pp-oc-corrs-11.eps}}
%the mean posterior prediction (grey point) and 95\% posterior
%prediction intervals (grey line segment) for Kendall's $\tau$ values
%calculated using replicated data. Kendall's $\tau$ values computed from
%the observed data are denoted by a black ``X''.}
%%

Table~\ref{tab:spfa-reg-coef} displays posterior summaries for the
regression parameters.
We saw little change in our estimate for the parameter of interest,
$\beta_5$, the coefficient for frontal WMH in adding additional factors.
Thus, our substantive conclusion regarding the association between an
individual's executive functioning and the volume of white matter
hyperintensities in the frontal region of the brain remains the same
whether we use the one-factor model or the better fitting bifactor
model. Based on our semiparametric latent variable model, we expect a
1SD increase in frontal white matter hyperintensity volume to be
associated with a 0.335SD decrease in the primary factor. In examining
the other coefficients, we see that none of the 95\% posterior credible
intervals have shifted to the extent that we would alter our posterior
belief about whether zero is a plausible value for the parameter.
However, the coefficients for sex, age and total brain volume did
decrease by 30--40\% in magnitude.

\section{Discussion}

In this paper we have developed a semiparametric latent variable model
for multivariate mixed outcome data. This model, unlike common
parametric latent variable modeling approaches for mixed outcome data
[\citet{sammel1997latent,moustaki2000generalized,dunson2003dynamic,shi1998bayesian}],
% Jonathan, please include the references for the common parametric
%approaches again
does not require the specification of conditional distributions for
each outcome given the latent variables.
When a data set combines a variety of mixed outcomes, picking
appropriate conditional distributions for each outcome encountered in
real data, extending the parametric models to account for all cases of
distributions, and extending estimation methods appropriately can be
labor-intensive.
%Analyzing multivariate cognitive test outcomes from a medical study,
%we found it labor-intensive to pick appropriate conditional
%distributions for each outcome encountered in real data, to extend the
%parametric models to account for all cases of distributions, and to
%extend estimation methods appropriately.
Moreover, specification of outcome conditional distributions given the
latent variables may be of little interest by itself in any research
setting where the main question is in the relationship between a common
factor (or factors) and a covariate of interest. Our proposed
semiparametric latent variable framework allows one to model
interdependencies among observed mixed outcome variables by specifying
an appropriate latent variable model while, at the same time, avoiding
the specification of outcome distributions conditional on the common
latent variables. We have demonstrated this approach for the
single-factor and bifactor models, incorporating a covariate effect on
the general factor.\looseness=-1

The extended rank likelihood can readily be employed with other latent
variable models, including item response theory models [\citet
{van1997handbook}] and structural equation models [\citet
{bollen1989structural}]. In structural equation models, the focus is
often on characterizing the relationship between latent variables
and/or between latent variables and fixed covariates as in the case of
our hierarchical model. In such cases where the focus is not on the
loadings or outcome-related parameters, the proposed semiparametric
approach would be quite useful in dealing with mixed outcome data.
However, the extended rank likelihood may not be as useful in cases
where outcome-specific parameters on the scale of the observed outcomes
are of interest. In item response theory models, one is often
interested in examining the item difficulty and discrimination
parameters to better understand the characteristics of individual test
questions. The difficulty parameter, the analogue\vadjust{\goodbreak} to the specific mean
in the factor model, is not directly identifiable with the extended
rank likelihood approach. Nonetheless, one could still carry out
posterior inference by relying on the relationship between the
difficulty parameter and the latent trait. For example, in a
two-parameter item response theory model for binary outcomes, the
probability of a positive response when the factor score is set to zero
is a one-to-one function of the difficulty parameter. Such an
alternative, however, may render the semiparametric approach less
convenient for a practitioner who is primarily interested in parameters
characterizing the properties of individual outcomes.

We employed the semiparametric latent variable model to study the
association between the volume of white matter hyperintensities in the
frontal lobe and cognitive testing outcomes related to executive
functioning from the Subcortical Ischemic Vascular Dementia (SIVD)
study. %White matter hyperintensities are areas of increased signal
%intensity in the brain.
The semiparametric latent variable model allowed us to analyze the
mixed cognitive testing outcomes without requiring the specification of
parametric distributions for the outcomes conditional on the latent variables.
It has been hypothesized that a greater volume of frontal lobe white
matter hyperintensities will be associated with worse executive functioning.
Consistent with this hypothesis, we found a negative association
between the primary factor in our model and the volume of white matter
hyperintensities.

Our model selection process was guided by substantive beliefs that
associations among items in the cognitive testing data are primarily
driven by the main latent factor but can potentially be influenced by
secondary latent factors due to local dependencies among groups of
related items. Thus,
we started our model-building process by fitting the one-factor
semiparametric model and relied on posterior predictive model checks to
evaluate model misfit and to guide us in identifying a secondary factor
structure for the bifactor model. Our posterior predictive checks
approach can therefore be thought of as a method of exploratory
bifactor analysis when the secondary factor structure is not known in
advance~[\citet{jennrich2011exploratory}]. It also provides a mechanism
by which statistical methodologists can work together with substantive
experts to develop models that are theoretically justified and that are
consistent with the data.
We note, however, that the main conclusion about the association
between executive functioning and regional brain volumes was not
affected much by the choice of a better fitting bifactor model over the
single factor model in our case.

While our proposed model selection process is somewhat ad-hoc, one
could explore the use of more formal model fit criteria, other model
selection methods or a fully Bayesian approach to determine the factor
structure for our semiparametric model. For example, one could use the
methods of \citet{knowles2011nonparametric} and \citet{rai2009infinite}
to incorporate the Indian Buffet Process prior to simultaneously
estimate the loadings, the loadings structure and the number of
factors. Within the bifactor model framework, \citet
{jennrich2011exploratory} recently proposed using a rotation criterion
to explore the secondary factor structure.\vadjust{\goodbreak} \citet{dunson2006efficient}
presented a Bayesian model averaging approach that accounts for the
uncertainty in the number of factors.

Overall, in our work with the cognitive testing data, we found that the
semiparametric model was more elegant and much easier in implementation
than the standard parametric approaches for mixed outcome data.
% Jonathan: I find this last few sentences a bit weak. It does not help
%to end the paper on a high note. In addition, it raises a new problem
%that only a few people would understand (in the way it is written
%here. I suggest nixing it. -- Elena

%Nonetheless, a formal comparison of the two methods, including an
%application of the two methods to the same data, is needed to fully
%understand the differences and impact on conclusions.
%In addition, an alternative approach to handling mixed outcomes in
%practice is to use mainstream latent variable modeling software and
%treat all outcomes as ordered categorical. In cases where the number
%of categories exceeds the number that can be accommodated by existing
%software, the outcomes are typically consolidated into fewer
%categories to satisfy the software's constraints. It would similarly
%be useful to understand how the semiparametric model performs relative
%to this software-constrained approach.

\section*{Acknowledgments}
The authors would like to thank Peter Hoff, S. McKay Curtis, Thomas
Richardson, Dan Mungas, Laura Gibbons and the Cognitive Outcomes with
Advanced Pyschometrics group, University of Washington, for many
helpful discussions and comments on earlier versions of this work. In
addition, the reviewers of the original submission provided valuable
feedback that have strengthened the final version.

% AOS,AOAS: If there are supplements please fill:
% \sname{Supplement A}
% \stitle{Additional Details Regarding Simulations}
% \slink[doi]{}
% \sdatatype{.pdf}
% \sdescription{The supplemental materials includes additional details
%regarding the simulations discussed in Section~\ref{sec:SIVD-app} and
%comparison of the estimation approach with a standard Gibbs sampling
%approach.}

%
% imsref loaded by akundreckaite, 2013-10-15 12:44:42
% imsref loaded by akundreckaite, 2013-10-15 13:24:31
% imsref loaded by akundreckaite, 2013-10-15 13:24:43
%

% zodis "Acknowledgments" paliekamas pagal autoriu

%suskaldyti doi

\printaddresses


\begin{thebibliography}{54}
% BibTex style file: ims.bst, 2013-01-28
% Default style options (sort=0,type=number).
% Used options (sort=1,type=nameyear).

%b1 #&#
\bibitem[\protect\citeauthoryear{Anderson}{2003}]{anderson2003introduction}
%
\begin{bbook}[mr]
\bauthor{\bsnm{Anderson},~\bfnm{T.~W.}\binits{T.~W.}}
(\byear{2003}).
\btitle{An Introduction to Multivariate Statistical Analysis},
\bedition{3rd} ed.
\bpublisher{Wiley}, \blocation{Hoboken, NJ}.
\bid{mr={1990662}}
\bptok{imsref}%
\end{bbook}
%
\endbibitem

%b2 #&#
\bibitem[\protect\citeauthoryear{Bartholomew, Knott and
Moustaki}{2011}]{bartholomew2011latent}
%
\begin{bbook}[mr]
\bauthor{\bsnm{Bartholomew},~\bfnm{David}\binits{D.}},
\bauthor{\bsnm{Knott},~\bfnm{Martin}\binits{M.}} \AND
\bauthor{\bsnm{Moustaki},~\bfnm{Irini}\binits{I.}}
(\byear{2011}).
\btitle{Latent Variable Models and Factor Analysis: A Unified Approach},
\bedition{3rd} ed.
\bpublisher{Wiley}, \blocation{Chichester}.
\bid{doi={10.1002/9781119970583}, mr={2849614}}
\bptok{imsref}%
\end{bbook}
%
\endbibitem

%b3 #&#
\bibitem[\protect\citeauthoryear{Bollen}{1989}]{bollen1989structural}
%
\begin{bbook}[mr]
\bauthor{\bsnm{Bollen},~\bfnm{Kenneth~A.}\binits{K.~A.}}
(\byear{1989}).
\btitle{Structural Equations with Latent Variables}.
\bpublisher{Wiley}, \blocation{New York}.
\bid{mr={0996025}}
\bptok{imsref}%
\end{bbook}
%
\endbibitem

%b4 #&#
\bibitem[\protect\citeauthoryear{Cardenas
et~al.}{2001}]{cardenas2001reliability}
%
\begin{barticle}[author]
\bauthor{\bsnm{Cardenas},~\bfnm{V.~A.}\binits{V.~A.}},
\bauthor{\bsnm{Ezekiel},~\bfnm{F.}\binits{F.}},
\bauthor{\bsnm{Di~Sclafani},~\bfnm{V.}\binits{V.}},
\bauthor{\bsnm{Gomberg},~\bfnm{B.}\binits{B.}} \AND
\bauthor{\bsnm{Fein},~\bfnm{G.}\binits{G.}}
(\byear{2001}).
\btitle{Reliability of tissue volumes and their spatial distribution for
segmented magnetic resonance images}.
\bjournal{Psychiatry Research: Neuroimaging}
\bvolume{106}
\bpages{193--205}.
\bptok{imsref}%
\end{barticle}
%
\endbibitem

%b5 #&#
\bibitem[\protect\citeauthoryear{Cattell}{1966}]{cattell1966scree}
%
\begin{barticle}[author]
\bauthor{\bsnm{Cattell},~\bfnm{R.~B.}\binits{R.~B.}}
(\byear{1966}).
\btitle{The scree test for the number of factors}.
\bjournal{Multivariate Behavioral Research}
\bvolume{1}
\bpages{245--276}.
\bptok{imsref}%
\end{barticle}
%
\endbibitem

%b6 #&#
\bibitem[\protect\citeauthoryear{Chui}{2007}]{chui2007subcortical}
%
\begin{barticle}[pbm]
\bauthor{\bsnm{Chui},~\bfnm{Helena~C.}\binits{H.~C.}}
(\byear{2007}).
\btitle{Subcortical ischemic vascular dementia}.
\bjournal{Neurol. Clin.}
\bvolume{25}
\bpages{717--740,~vi}.
\bid{doi={10.1016/j.ncl.2007.04.003}, issn={0733-8619}, mid={NIHMS28661},
pii={S0733-8619(07)00059-X}, pmcid={2084201}, pmid={17659187}}
\bptok{imsref}%
\end{barticle}
%
\endbibitem

%b7 #&#
\bibitem[\protect\citeauthoryear{Chui et~al.}{2006}]{chui2006cognitive}
%
\begin{barticle}[author]
\bauthor{\bsnm{Chui},~\bfnm{H.~C.}\binits{H.~C.}},
\bauthor{\bsnm{Zarow},~\bfnm{C.}\binits{C.}},
\bauthor{\bsnm{Mack},~\bfnm{W.~J.}\binits{W.~J.}},
\bauthor{\bsnm{Ellis},~\bfnm{W.~G.}\binits{W.~G.}},
\bauthor{\bsnm{Zheng},~\bfnm{L.}\binits{L.}},
\bauthor{\bsnm{Jagust},~\bfnm{W.~J.}\binits{W.~J.}},
\bauthor{\bsnm{Mungas},~\bfnm{D.}\binits{D.}},
\bauthor{\bsnm{Reed},~\bfnm{B.~R.}\binits{B.~R.}},
\bauthor{\bsnm{Kramer},~\bfnm{J.~H.}\binits{J.~H.}},
\bauthor{\bsnm{DeCarli},~\bfnm{C.~C.}\binits{C.~C.}} \betal{et~al.}
(\byear{2006}).
\btitle{{Cognitive impact of subcortical vascular and Alzheimer's disease
pathology}}.
\bjournal{Annals of Neurology}
\bvolume{60}
\bpages{677}.
\bptok{imsref}%
\end{barticle}
%
\endbibitem

%b8 #&#
\bibitem[\protect\citeauthoryear{Congdon}{2003}]{congdon2003applied}
%
\begin{bbook}[mr]
\bauthor{\bsnm{Congdon},~\bfnm{Peter}\binits{P.}}
(\byear{2003}).
\btitle{Applied {B}ayesian Modelling}.
\bpublisher{Wiley}, \blocation{Chichester}.
\bid{doi={10.1002/0470867159}, mr={1990543}}
\bptok{imsref}%
\end{bbook}
%
\endbibitem

%b9 #&#
\bibitem[\protect\citeauthoryear{Congdon}{2006}]{congdon2006bayesian}
%
\begin{bbook}[mr]
\bauthor{\bsnm{Congdon},~\bfnm{Peter}\binits{P.}}
(\byear{2006}).
\btitle{Bayesian Statistical Modelling},
\bedition{2nd} ed.
\bpublisher{Wiley}, \blocation{Chichester}.
\bid{doi={10.1002/9780470035948}, mr={2281386}}
\bptok{imsref}%
\end{bbook}
%
\endbibitem

%b10 #&#
\bibitem[\protect\citeauthoryear{Dobra and Lenkoski}{2011}]{dobra2011copula}
%
\begin{barticle}[mr]
\bauthor{\bsnm{Dobra},~\bfnm{Adrian}\binits{A.}} \AND
\bauthor{\bsnm{Lenkoski},~\bfnm{Alex}\binits{A.}}
(\byear{2011}).
\btitle{Copula {G}aussian graphical models and their application to modeling
functional disability data}.
\bjournal{Ann. Appl. Stat.}
\bvolume{5}
\bpages{969--993}.
\bid{doi={10.1214/10-AOAS397}, issn={1932-6157}, mr={2840183}}
\bptok{imsref}%
\end{barticle}
%
\endbibitem

%b11 #&#
\bibitem[\protect\citeauthoryear{Dunn}{1973}]{dunn1973note}
%
\begin{barticle}[mr]
\bauthor{\bsnm{Dunn},~\bfnm{James~E.}\binits{J.~E.}}
(\byear{1973}).
\btitle{A note on a sufficiency condition for uniqueness of restricted factor
matrix}.
\bjournal{Psychometrika}
\bvolume{38}
\bpages{141--143}.
\bid{issn={0033-3123}, mr={0345326}}
\bptok{imsref}%
\end{barticle}
%
\endbibitem

%b12 #&#
\bibitem[\protect\citeauthoryear{Dunson}{2003}]{dunson2003dynamic}
%
\begin{barticle}[mr]
\bauthor{\bsnm{Dunson},~\bfnm{David~B.}\binits{D.~B.}}
(\byear{2003}).
\btitle{Dynamic latent trait models for multidimensional longitudinal data}.
\bjournal{J. Amer. Statist. Assoc.}
\bvolume{98}
\bpages{555--563}.
\bid{doi={10.1198/016214503000000387}, issn={0162-1459}, mr={2011671}}
\bptok{imsref}%
\end{barticle}
%
\endbibitem

%b13 #&#
\bibitem[\protect\citeauthoryear{Dunson et~al.}{2006}]{dunson2006efficient}
%
\begin{bmisc}[author]
\bauthor{\bsnm{Dunson},~\bfnm{D.~B.}\binits{D.~B.}} \betal{et~al.}
(\byear{2006}).
\bhowpublished{Efficient Bayesian model averaging in factor analysis. Technical
report, Duke Univ., Durham, NC.}
\bptok{imsref}%
\end{bmisc}
%
\endbibitem

%b14 #&#
\bibitem[\protect\citeauthoryear{Erosheva and
Curtis}{2011}]{erosheva2011specification}
%
\begin{bmisc}[author]
\bauthor{\bsnm{Erosheva},~\bfnm{E.}\binits{E.}} \AND
\bauthor{\bsnm{Curtis},~\bfnm{S.~M.}\binits{S.~M.}}
(\byear{2011}).
\bhowpublished{Specification of rotational constraints in Bayesian confirmatory
factor analysis. Technical Report No. 589, Univ. Washington, Seattle, WA.}
\bptok{imsref}%
\end{bmisc}
%
\endbibitem

%b15 #&#
\bibitem[\protect\citeauthoryear{Geweke}{1992}]{geweke1992evaluating}
%
\begin{bincollection}[mr]
\bauthor{\bsnm{Geweke},~\bfnm{John}\binits{J.}}
(\byear{1992}).
\btitle{Evaluating the accuracy of sampling-based approaches to the calculation
of posterior moments}.
In \bbooktitle{Bayesian Statistics, 4 ({P}e\~n\'\i Scola, 1991)}
(\beditor{\bfnm{J.~M.}\binits{J.~M.}~\bsnm{Bernardo}},
\beditor{\bfnm{J.}\binits{J.}~\bsnm{Berger}},
\beditor{\bfnm{A.~P.}\binits{A.~P.}~\bsnm{Dawid}} \AND
\beditor{\bfnm{J.~F.~M.}\binits{J.~F.~M.}~\bsnm{Smith}}, eds.)
\bpages{169--193}.
\bpublisher{Oxford Univ. Press}, \blocation{New York}.
\bid{mr={1380276}}
\bptok{imsref}%
\end{bincollection}\vadjust{\goodbreak}
%
\endbibitem

%b16 #&#
\bibitem[\protect\citeauthoryear{Geweke and Zhou}{1996}]{geweke1996measuring}
%
\begin{barticle}[author]
\bauthor{\bsnm{Geweke},~\bfnm{J.}\binits{J.}} \AND
\bauthor{\bsnm{Zhou},~\bfnm{G.}\binits{G.}}
(\byear{1996}).
\btitle{Measuring the pricing error of the arbitrage pricing theory}.
\bjournal{Review of Financial Studies}
\bvolume{9}
\bpages{557--587}.
\bptok{imsref}%
\end{barticle}
%
\endbibitem

%b17 #&#
\bibitem[\protect\citeauthoryear{Ghosh and Dunson}{2008}]{ghosh2008bayesian}
%
\begin{bincollection}[author]
\bauthor{\bsnm{Ghosh},~\bfnm{J.}\binits{J.}} \AND
\bauthor{\bsnm{Dunson},~\bfnm{D.~B.}\binits{D.~B.}}
(\byear{2008}).
\btitle{Bayesian model selection in factor analytic models}.
In \bbooktitle{Random effect and latent variable model selection}
\bpages{151--163}.
\bpublisher{Springer}, \blocation{New York}.
\bptok{imsref}%
\end{bincollection}
%
\endbibitem

%b18 #&#
\bibitem[\protect\citeauthoryear{Ghosh and Dunson}{2009}]{ghosh2009default}
%
\begin{barticle}[mr]
\bauthor{\bsnm{Ghosh},~\bfnm{Joyee}\binits{J.}} \AND
\bauthor{\bsnm{Dunson},~\bfnm{David~B.}\binits{D.~B.}}
(\byear{2009}).
\btitle{Default prior distributions and efficient posterior
computation in
{B}ayesian factor analysis}.
\bjournal{J. Comput. Graph. Statist.}
\bvolume{18}
\bpages{306--320}.
\bid{doi={10.1198/jcgs.2009.07145}, issn={1061-8600}, mr={2749834}}
\bptok{imsref}%
\end{barticle}
%
\endbibitem

%b19 #&#
\bibitem[\protect\citeauthoryear{Gruhl, Erosheva and
Crane}{2010}]{gruhl2010analyzing}
%
\begin{bmisc}[author]
\bauthor{\bsnm{Gruhl},~\bfnm{J.}\binits{J.}},
\bauthor{\bsnm{Erosheva},~\bfnm{E.}\binits{E.}} \AND
\bauthor{\bsnm{Crane},~\bfnm{P.}\binits{P.}}
(\byear{2010}).
\bhowpublished{Analyzing cognitive testing data with extensions of item
response theory models. Presented at the Joint Statistical Meetings,
Vancouver, Canada, August 3, 2010}.
\bptok{imsref}%
\end{bmisc}
%
\endbibitem

%b20 #&#
\bibitem[\protect\citeauthoryear{Gruhl, Erosheva and
Crane}{2011}]{gruhl2011semiparametric}
%
\begin{bmisc}[author]
\bauthor{\bsnm{Gruhl},~\bfnm{J.}\binits{J.}},
\bauthor{\bsnm{Erosheva},~\bfnm{E.}\binits{E.}} \AND
\bauthor{\bsnm{Crane},~\bfnm{P.}\binits{P.}}
(\byear{2011}).
\bhowpublished{A semiparametric Bayesian latent trait model for multivariate
mixed type data. In \textit{International Meeting of the Pyschometric Society}}.
\bptok{imsref}%
\end{bmisc}
%
\endbibitem

%b21 #&#
\bibitem[\protect\citeauthoryear{Guttman}{1954}]{guttman1954some}
%
\begin{barticle}[mr]
\bauthor{\bsnm{Guttman},~\bfnm{Louis}\binits{L.}}
(\byear{1954}).
\btitle{Some necessary conditions for common-factor analysis}.
\bjournal{Psychometrika}
\bvolume{19}
\bpages{149--161}.
\bid{issn={0033-3123}, mr={0091235}}
\bptok{imsref}%
\end{barticle}
%
\endbibitem

%b22 #&#
\bibitem[\protect\citeauthoryear{Hachinski
et~al.}{2006}]{hachinski2006national}
%
\begin{barticle}[author]
\bauthor{\bsnm{Hachinski},~\bfnm{V.}\binits{V.}},
\bauthor{\bsnm{Iadecola},~\bfnm{C.}\binits{C.}},
\bauthor{\bsnm{Petersen},~\bfnm{R.~C.}\binits{R.~C.}},
\bauthor{\bsnm{Breteler},~\bfnm{M.~M.}\binits{M.~M.}},
\bauthor{\bsnm{Nyenhuis},~\bfnm{D.~L.}\binits{D.~L.}},
\bauthor{\bsnm{Black},~\bfnm{S.~E.}\binits{S.~E.}},
\bauthor{\bsnm{Powers},~\bfnm{W.~J.}\binits{W.~J.}},
\bauthor{\bsnm{DeCarli},~\bfnm{C.}\binits{C.}},
\bauthor{\bsnm{Merino},~\bfnm{J.~G.}\binits{J.~G.}},
\bauthor{\bsnm{Kalaria},~\bfnm{R.~N.}\binits{R.~N.}} \betal{et~al.}
(\byear{2006}).
\btitle{National institute of neurological disorders and stroke---Canadian
stroke network vascular cognitive impairment harmonization standards}.
\bjournal{Stroke}
\bvolume{37}
\bpages{2220--2241}.
\bptok{imsref}%
\end{barticle}
%
\endbibitem

%b23 #&#
\bibitem[\protect\citeauthoryear{Hoff}{2007}]{hoff2007extending}
%
\begin{barticle}[mr]
\bauthor{\bsnm{Hoff},~\bfnm{Peter~D.}\binits{P.~D.}}
(\byear{2007}).
\btitle{Extending the rank likelihood for semiparametric copula estimation}.
\bjournal{Ann. Appl. Stat.}
\bvolume{1}
\bpages{265--283}.
\bid{doi={10.1214/07-AOAS107}, issn={1932-6157}, mr={2393851}}
\bptok{imsref}%
\end{barticle}
%
\endbibitem

%b24 #&#
\bibitem[\protect\citeauthoryear{Hoff}{2009}]{hoff2009first}
%
\begin{bbook}[mr]
\bauthor{\bsnm{Hoff},~\bfnm{Peter~D.}\binits{P.~D.}}
(\byear{2009}).
\btitle{A First Course in {B}ayesian Statistical Methods}.
\bpublisher{Springer}, \blocation{New York}.
\bid{doi={10.1007/978-0-387-92407-6}, mr={2648134}}
\bptok{imsref}%
\end{bbook}
%
\endbibitem

%b25 #&#
\bibitem[\protect\citeauthoryear{Holzinger and
Swineford}{1937}]{holzinger1937bi}
%
\begin{barticle}[author]
\bauthor{\bsnm{Holzinger},~\bfnm{K.~J.}\binits{K.~J.}} \AND
\bauthor{\bsnm{Swineford},~\bfnm{F.}\binits{F.}}
(\byear{1937}).
\btitle{The bi-factor method}.
\bjournal{Psychometrika}
\bvolume{2}
\bpages{41--54}.
\bptok{imsref}%
\end{barticle}
%
\endbibitem

%b26 #&#
\bibitem[\protect\citeauthoryear{Jennrich}{1978}]{jennrich1978rotational}
%
\begin{barticle}[mr]
\bauthor{\bsnm{Jennrich},~\bfnm{Robert~I.}\binits{R.~I.}}
(\byear{1978}).
\btitle{Rotational equivalence of factor loading matrices with specified
values}.
\bjournal{Psychometrika}
\bvolume{43}
\bpages{421--426}.
\bid{doi={10.1007/BF02293650}, issn={0033-3123}, mr={0514727}}
\bptok{imsref}%
\end{barticle}
%
\endbibitem

%b27 #&#
\bibitem[\protect\citeauthoryear{Jennrich and
Bentler}{2011}]{jennrich2011exploratory}
%
\begin{barticle}[mr]
\bauthor{\bsnm{Jennrich},~\bfnm{Robert~I.}\binits{R.~I.}} \AND
\bauthor{\bsnm{Bentler},~\bfnm{Peter~M.}\binits{P.~M.}}
(\byear{2011}).
\btitle{Exploratory bi-factor analysis}.
\bjournal{Psychometrika}
\bvolume{76}
\bpages{537--549}.
\bid{doi={10.1007/s11336-011-9218-4}, issn={0033-3123}, mr={2851500}}
\bptok{imsref}%
\end{barticle}
%
\endbibitem

%b28 #&#
\bibitem[\protect\citeauthoryear{J{\"o}reskog}{1969}]{joreskog1969general}
%
\begin{barticle}[author]
\bauthor{\bsnm{J{\"o}reskog},~\bfnm{K.~G.}\binits{K.~G.}}
(\byear{1969}).
\btitle{A general approach to confirmatory maximum likelihood factor analysis}.
\bjournal{Psychometrika}
\bvolume{34}
\bpages{183--202}.
\bptok{imsref}%
\end{barticle}
%
\endbibitem

%b29 #&#
\bibitem[\protect\citeauthoryear{Kl{\"u}ppelberg and
Kuhn}{2009}]{kluppelberg2009copula}
%
\begin{barticle}[mr]
\bauthor{\bsnm{Kl{\"u}ppelberg},~\bfnm{Claudia}\binits{C.}} \AND
\bauthor{\bsnm{Kuhn},~\bfnm{Gabriel}\binits{G.}}
(\byear{2009}).
\btitle{Copula structure analysis}.
\bjournal{J. R. Stat. Soc. Ser. B Stat. Methodol.}
\bvolume{71}
\bpages{737--753}.
\bid{doi={10.1111/j.1467-9868.2009.00707.x}, issn={1369-7412}, mr={2749917}}
\bptok{imsref}%
\end{barticle}
%
\endbibitem

%b30 #&#
\bibitem[\protect\citeauthoryear{Knowles and
Ghahramani}{2011}]{knowles2011nonparametric}
%
\begin{barticle}[mr]
\bauthor{\bsnm{Knowles},~\bfnm{David}\binits{D.}} \AND
\bauthor{\bsnm{Ghahramani},~\bfnm{Zoubin}\binits{Z.}}
(\byear{2011}).
\btitle{Nonparametric {B}ayesian sparse factor models with application
to gene
expression modeling}.
\bjournal{Ann. Appl. Stat.}
\bvolume{5}
\bpages{1534--1552}.
\bid{doi={10.1214/10-AOAS435}, issn={1932-6157}, mr={2849785}}
\bptok{imsref}%
\end{barticle}
%
\endbibitem

%b31 #&#
\bibitem[\protect\citeauthoryear{Kuczynski et~al.}{2010}]{kuczynski2010white}
%
\begin{barticle}[author]
\bauthor{\bsnm{Kuczynski},~\bfnm{B.}\binits{B.}},
\bauthor{\bsnm{Targan},~\bfnm{E.}\binits{E.}},
\bauthor{\bsnm{Madison},~\bfnm{C.}\binits{C.}},
\bauthor{\bsnm{Weiner},~\bfnm{M.}\binits{M.}},
\bauthor{\bsnm{Zhang},~\bfnm{Y.}\binits{Y.}},
\bauthor{\bsnm{Reed},~\bfnm{B.}\binits{B.}},
\bauthor{\bsnm{Chui},~\bfnm{H.~C.}\binits{H.~C.}} \AND
\bauthor{\bsnm{Jagust},~\bfnm{W.}\binits{W.}}
(\byear{2010}).
\btitle{White matter integrity and cortical metabolic associations in
aging and
dementia}.
\bjournal{Alzheimer's and Dementia}
\bvolume{6}
\bpages{54--62}.
\bptok{imsref}%
\end{barticle}
%
\endbibitem

%b32 #&#
\bibitem[\protect\citeauthoryear{Liu, Rubin and Wu}{1998}]{liu1998parameter}
%
\begin{barticle}[mr]
\bauthor{\bsnm{Liu},~\bfnm{Chuanhai}\binits{C.}},
\bauthor{\bsnm{Rubin},~\bfnm{Donald~B.}\binits{D.~B.}} \AND
\bauthor{\bsnm{Wu},~\bfnm{Ying~Nian}\binits{Y.~N.}}
(\byear{1998}).
\btitle{Parameter expansion to accelerate {EM}: The {PX}-{EM} algorithm}.
\bjournal{Biometrika}
\bvolume{85}
\bpages{755--770}.
\bid{doi={10.1093/biomet/85.4.755}, issn={0006-3444}, mr={1666758}}
\bptok{imsref}%
\end{barticle}
%
\endbibitem

%b33 #&#
\bibitem[\protect\citeauthoryear{Liu and Wu}{1999}]{liu1999parameter}
%
\begin{barticle}[mr]
\bauthor{\bsnm{Liu},~\bfnm{Jun~S.}\binits{J.~S.}} \AND
\bauthor{\bsnm{Wu},~\bfnm{Ying~Nian}\binits{Y.~N.}}
(\byear{1999}).
\btitle{Parameter expansion for data augmentation}.
\bjournal{J. Amer. Statist. Assoc.}
\bvolume{94}
\bpages{1264--1274}.
\bid{doi={10.2307/2669940}, issn={0162-1459}, mr={1731488}}
\bptok{imsref}%
\end{barticle}
%
\endbibitem

%b34 #&#
\bibitem[\protect\citeauthoryear{Loken}{2005}]{loken2005identification}
%
\begin{barticle}[mr]
\bauthor{\bsnm{Loken},~\bfnm{Eric}\binits{E.}}
(\byear{2005}).
\btitle{Identification constraints and inference in factor models}.
\bjournal{Struct. Equ. Model.}
\bvolume{12}
\bpages{232--244}.
\bid{doi={10.1207/s15328007sem1202_3}, issn={1070-5511}, mr={2135915}}
\bptok{imsref}%
\end{barticle}
%
\endbibitem

%b35 #&#
\bibitem[\protect\citeauthoryear{Lopes and West}{2004}]{lopes2004bayesian}
%
\begin{barticle}[mr]
\bauthor{\bsnm{Lopes},~\bfnm{Hedibert~Freitas}\binits{H.~F.}} \AND
\bauthor{\bsnm{West},~\bfnm{Mike}\binits{M.}}
(\byear{2004}).
\btitle{Bayesian model assessment in factor analysis}.
\bjournal{Statist. Sinica}
\bvolume{14}
\bpages{41--67}.
\bid{issn={1017-0405}, mr={2036762}}
\bptok{imsref}%
\end{barticle}
%
\endbibitem

%b36 #&#
\bibitem[\protect\citeauthoryear{Millsap}{2001}]{millsap2001trivial}
%
\begin{barticle}[author]
\bauthor{\bsnm{Millsap},~\bfnm{R.~E.}\binits{R.~E.}}
(\byear{2001}).
\btitle{When trivial constraints are not trivial: The choice of uniqueness
constraints in confirmatory factor analysis}.
\bjournal{Struct. Equ. Model.}
\bvolume{8}
\bpages{1--17}.
\bptok{imsref}%
\end{barticle}
%
\endbibitem

%b37 #&#
\bibitem[\protect\citeauthoryear{Morris}{1993}]{morris1993clinical}
%
\begin{barticle}[pbm]
\bauthor{\bsnm{Morris},~\bfnm{J.~C.}\binits{J.~C.}}
(\byear{1993}).
\btitle{The Clinical Dementia Rating (CDR): Current version and
scoring rules}.
\bjournal{Neurology}
\bvolume{43}
\bpages{2412--2414}.
\bid{issn={0028-3878}, pmid={8232972}}
\bptok{imsref}%
\end{barticle}
%
\endbibitem

%b38 #&#
\bibitem[\protect\citeauthoryear{Morris}{1997}]{morris1997clinical}
%
\begin{barticle}[pbm]
\bauthor{\bsnm{Morris},~\bfnm{J.~C.}\binits{J.~C.}}
(\byear{1997}).
\btitle{Clinical dementia rating: A reliable and valid diagnostic and staging
measure for dementia of the Alzheimer type}.
\bjournal{Int. Psychogeriatr.}
\bvolume{9 Suppl 1}
\bpages{173--176; discussion 177--178}.
\bid{issn={1041-6102}, pmid={9447441}}
\bptnote{check related}%
\bptok{imsref}%
\end{barticle}
%
\endbibitem

%b39 #&#
\bibitem[\protect\citeauthoryear{Moustaki and
Knott}{2000}]{moustaki2000generalized}
%
\begin{barticle}[mr]
\bauthor{\bsnm{Moustaki},~\bfnm{Irini}\binits{I.}} \AND
\bauthor{\bsnm{Knott},~\bfnm{Martin}\binits{M.}}
(\byear{2000}).
\btitle{Generalized latent trait models}.
\bjournal{Psychometrika}
\bvolume{65}
\bpages{391--411}.
\bid{doi={10.1007/BF02296153}, issn={0033-3123}, mr={1792703}}
\bptok{imsref}%
\end{barticle}
%
\endbibitem

%b40 #&#
\bibitem[\protect\citeauthoryear{Mungas
et~al.}{2005}]{mungas2005longitudinal}
%
\begin{barticle}[author]
\bauthor{\bsnm{Mungas},~\bfnm{D.}\binits{D.}},
\bauthor{\bsnm{Harvey},~\bfnm{D.}\binits{D.}},
\bauthor{\bsnm{Reed},~\bfnm{B.~R.}\binits{B.~R.}},
\bauthor{\bsnm{Jagust},~\bfnm{W.~J.}\binits{W.~J.}},
\bauthor{\bsnm{DeCarli},~\bfnm{C.}\binits{C.}},
\bauthor{\bsnm{Beckett},~\bfnm{L.}\binits{L.}},
\bauthor{\bsnm{Mack},~\bfnm{W.~J.}\binits{W.~J.}},
\bauthor{\bsnm{Kramer},~\bfnm{J.~H.}\binits{J.~H.}},
\bauthor{\bsnm{Weiner},~\bfnm{M.~W.}\binits{M.~W.}},
\bauthor{\bsnm{Schuff},~\bfnm{N.}\binits{N.}} \betal{et~al.}
(\byear{2005}).
\btitle{Longitudinal volumetric MRI change and rate of cognitive decline}.
\bjournal{Neurology}
\bvolume{65}
\bpages{565--571}.
\bptok{imsref}%
\end{barticle}
%
\endbibitem

%b41 #&#
\bibitem[\protect\citeauthoryear{Muraki}{1992}]{muraki1992generalized}
%
\begin{barticle}[author]
\bauthor{\bsnm{Muraki},~\bfnm{E.}\binits{E.}}
(\byear{1992}).
\btitle{{A generalized partial credit model: Application of an EM algorithm}}.
\bjournal{Appl. Psychol. Meas.}
\bvolume{16}
\bpages{159}.
\bptok{imsref}%
\end{barticle}
%
\endbibitem

%b42 #&#
\bibitem[\protect\citeauthoryear{Murray et~al.}{2013}]{murray2011bayesian}
%
\begin{barticle}[author]
\bauthor{\bsnm{Murray},~\bfnm{Jared~S.}\binits{J.~S.}},
\bauthor{\bsnm{Dunson},~\bfnm{David~B.}\binits{D.~B.}},
\bauthor{\bsnm{Carin},~\bfnm{Lawrence}\binits{L.}} \AND
\bauthor{\bsnm{Lucas},~\bfnm{Joseph~E.}\binits{J.~E.}}
(\byear{2013}).
\btitle{Bayesian Gaussian copula factor models for mixed data}.
\bjournal{J. Amer. Statist. Assoc.}
\bvolume{108}
\bpages{656--665}.
\bptok{imsref}%
\end{barticle}
%
\endbibitem

%b43 #&#
\bibitem[\protect\citeauthoryear{Pettitt}{1982}]{pettitt1982inference}
%
\begin{barticle}[mr]
\bauthor{\bsnm{Pettitt},~\bfnm{A.~N.}\binits{A.~N.}}
(\byear{1982}).
\btitle{Inference for the linear model using a likelihood based on ranks}.
\bjournal{J. R. Stat. Soc. Ser. B Stat. Methodol.}
\bvolume{44}
\bpages{234--243}.
\bid{issn={0035-9246}, mr={0676214}}
\bptok{imsref}%
\end{barticle}
%
\endbibitem

%b44 #&#
\bibitem[\protect\citeauthoryear{Raftery and Lewis}{1995}]{raftery1995number}
%
\begin{bincollection}[author]
\bauthor{\bsnm{Raftery},~\bfnm{A.~E.}\binits{A.~E.}} \AND
\bauthor{\bsnm{Lewis},~\bfnm{S.~M.}\binits{S.~M.}}
(\byear{1995}).
\btitle{{The number of iterations, convergence diagnostics and generic
Metropolis algorithms}}.
In \bbooktitle{Practical Markov Chain Monte Carlo}
(\beditor{\bfnm{W.~R.}\binits{W.~R.}~\bsnm{Gilks}},
\beditor{\bfnm{D.~J.}\binits{D.~J.}~\bsnm{Spiegelhalter}} \AND
\beditor{\bfnm{S.}\binits{S.}~\bsnm{Richardson}}, eds.).
\bpublisher{Chapman \& Hall}, \blocation{London, UK.}
\bptok{imsref}%
\end{bincollection}
%
\endbibitem

%b45 #&#
\bibitem[\protect\citeauthoryear{Rai and Daum{\'
e}~III}{2009}]{rai2009infinite}
%
\begin{bmisc}[author]
\bauthor{\bsnm{Rai},~\bfnm{P.}\binits{P.}} \AND
\bauthor{\bsnm{Daum{\'e}~III},~\bfnm{H.}\binits{H.}}
(\byear{2009}).
\bhowpublished{The infinite hierarchical factor regression model.
Available at
\arxivurl{arXiv:0908.0570}.}
\bptok{imsref}%
\end{bmisc}
%
\endbibitem

%b46 #&#
\bibitem[\protect\citeauthoryear{Reise, Morizot and
Hays}{2007}]{reise2007role}
%
\begin{barticle}[pbm]
\bauthor{\bsnm{Reise},~\bfnm{Steven~P.}\binits{S.~P.}},
\bauthor{\bsnm{Morizot},~\bfnm{Julien}\binits{J.}} \AND
\bauthor{\bsnm{Hays},~\bfnm{Ron~D.}\binits{R.~D.}}
(\byear{2007}).
\btitle{The role of the bifactor model in resolving dimensionality
issues in
health outcomes measures}.
\bjournal{Qual. Life Res.}
\bvolume{16 Suppl 1}
\bpages{19--31}.
\bid{doi={10.1007/s11136-007-9183-7}, issn={0962-9343}, pmid={17479357}}
\bptok{imsref}%
\end{barticle}
%
\endbibitem

%b47 #&#
\bibitem[\protect\citeauthoryear{Samejima}{1969}]{samejima1969estimation}
%
\begin{barticle}[author]
\bauthor{\bsnm{Samejima},~\bfnm{F.}\binits{F.}}
(\byear{1969}).
\btitle{Estimation of latent ability using a response pattern of graded
scores.}
\bjournal{Psychometrika Monograph Supplement}
\bvolume{34}
\bpages{1--100}.
\bptok{imsref}%
\end{barticle}
%
\endbibitem

%b48 #&#
\bibitem[\protect\citeauthoryear{Sammel, Ryan and
Legler}{1997}]{sammel1997latent}
%
\begin{barticle}[author]
\bauthor{\bsnm{Sammel},~\bfnm{M.~D.}\binits{M.~D.}},
\bauthor{\bsnm{Ryan},~\bfnm{L.~M.}\binits{L.~M.}} \AND
\bauthor{\bsnm{Legler},~\bfnm{J.~M.}\binits{J.~M.}}
(\byear{1997}).
\btitle{{Latent variable models for mixed discrete and continuous outcomes}}.
\bjournal{J. R. Stat. Soc. Ser. B Stat. Methodol.}
\bvolume{59}
\bpages{667--678}.
\bptok{imsref}%
\end{barticle}
%
\endbibitem

%b49 #&#
\bibitem[\protect\citeauthoryear{Shi and Lee}{1998}]{shi1998bayesian}
%
\begin{barticle}[author]
\bauthor{\bsnm{Shi},~\bfnm{J.~Q.}\binits{J.~Q.}} \AND
\bauthor{\bsnm{Lee},~\bfnm{S.~Y.}\binits{S.~Y.}}
(\byear{1998}).
\btitle{{Bayesian sampling-based approach for factor analysis models with
continuous and polytomous data}}.
\bjournal{British J. Math. Statist. Psych.}
\bvolume{51}
\bpages{233--252}.
\bptok{imsref}%
\end{barticle}
%
\endbibitem

%b50 #&#
\bibitem[\protect\citeauthoryear{Skrondal and
Rabe-Hesketh}{2004}]{skrondal2004generalized}
%
\begin{bbook}[mr]
\bauthor{\bsnm{Skrondal},~\bfnm{Anders}\binits{A.}} \AND
\bauthor{\bsnm{Rabe-Hesketh},~\bfnm{Sophia}\binits{S.}}
(\byear{2004}).
\btitle{Generalized Latent Variable Modeling: Multilevel,
Longitudinal, and
Structural Equation Models}.
\bpublisher{Chapman \& Hall/CRC}, \blocation{Boca Raton, FL}.
\bid{doi={10.1201/9780203489437}, mr={2059021}}
\bptok{imsref}%
\end{bbook}
%
\endbibitem

%b51 #&#
\bibitem[\protect\citeauthoryear{Stephens}{2000}]{stephens2000dealing}
%
\begin{barticle}[mr]
\bauthor{\bsnm{Stephens},~\bfnm{Matthew}\binits{M.}}
(\byear{2000}).
\btitle{Dealing with label switching in mixture models}.
\bjournal{J. R. Stat. Soc. Ser. B Stat. Methodol.}
\bvolume{62}
\bpages{795--809}.
\bid{doi={10.1111/1467-9868.00265}, issn={1369-7412}, mr={1796293}}
\bptok{imsref}%
\end{barticle}
%
\endbibitem

%b52 #&#
\bibitem[\protect\citeauthoryear{van~der Linden and
Hambleton}{1997}]{van1997handbook}
%
\begin{bbook}[mr]
\beditor{\bparticle{van~der} \bsnm{Linden},~\bfnm{Wim~J.}\binits
{W.~J.}} \AND
\beditor{\bsnm{Hambleton},~\bfnm{Ronald~K.}\binits{R.~K.}}, eds.
(\byear{1997}).
\btitle{Handbook of Modern Item Response Theory}.
\bpublisher{Springer}, \blocation{New York}.
\bid{mr={1601043}}
\bptok{imsref}%
\end{bbook}
%
\endbibitem

%b53 #&#
\bibitem[\protect\citeauthoryear{West}{1987}]{west1987scale}
%
\begin{barticle}[mr]
\bauthor{\bsnm{West},~\bfnm{Mike}\binits{M.}}
(\byear{1987}).
\btitle{On scale mixtures of normal distributions}.
\bjournal{Biometrika}
\bvolume{74}
\bpages{646--648}.
\bid{doi={10.1093/biomet/74.3.646}, issn={0006-3444}, mr={0909372}}
\bptok{imsref}%
\end{barticle}
%
\endbibitem

\end{thebibliography}
\end{document}